\documentclass[english,twocolumn,journal]{IEEEtran}
\usepackage[T1]{fontenc}
\usepackage{float}
\usepackage{amsthm}
\usepackage{amsmath}
\usepackage{amssymb}
\usepackage{graphicx}
\makeatletter
\floatstyle{ruled}
\newfloat{algorithm}{tbp}{loa}
\providecommand{\algorithmname}{Algorithm}
\floatname{algorithm}{\protect\algorithmname}
\theoremstyle{plain}

\theoremstyle{remark}

\IEEEoverridecommandlockouts
\usepackage{ifpdf}
\usepackage{cite}
\usepackage{amsmath}
\usepackage{float}
\usepackage{array}
\usepackage{makecell}
\usepackage{url}
\usepackage[outerbars,color]{changebar}
\cbcolor{red}
\makeatother
\usepackage{babel}
\providecommand{\remarkname}{Remark}
\providecommand{\theoremname}{Theorem}

\begin{document}
\title{ A Belief Propagation  Based  Framework for Soft Multiple-Symbol Differential Detection
}
\author{Chanfei Wang, Tiejun Lv, {\em Senior Member,~IEEE}, Hui Gao, {\em Member,~IEEE},\\
and Shaoshi Yang, \emph{Member, IEEE}
\thanks
{This work is supported by the National Natural Science Foundation of China (NSFC) (Grant No. 61271188 and  61401041), the Fundamental Research Funds for the Central Universities (Grant No. 2014RC0106), and the National High Technology Research and Development Program of China (863 Program)
(Grant No. 2015AA01A706). (\emph{Corresponding author: Shaoshi Yang})}
\thanks{C. Wang,  T. Lv, and H. Gao are  with the School of Information and Communication Engineering,
Beijing University of Posts and Telecommunications (BUPT), Beijing, China 100876 (email: \{wangchanfei,  lvtiejun, huigao\}@bupt.edu.cn).%
} %
\thanks{S. Yang is  with the School of Electronics and Computer Science, University
of Southampton, Southampton, SO17 1BJ, U.K.  (email: sy7g09@ecs.soton.ac.uk).%
}}

\maketitle

\begin{abstract}

Soft noncoherent  detection,  which relies on calculating the  \textit{a posteriori} probabilities (APPs) of the bits transmitted with no channel estimation,
is imperative for achieving excellent detection performance  in high-dimensional wireless communications.
In this paper, a  high-performance  belief propagation (BP)-based soft   multiple-symbol differential detection (MSDD)
framework, dubbed BP-MSDD,  is proposed  with its illustrative application in  differential space-time block-code (DSTBC)-aided  ultra-wideband impulse radio (UWB-IR) systems.
Firstly,  we revisit the  signal sampling with  the aid of  a trellis structure  and  decompose the trellis  into multiple subtrellises.
Furthermore, we derive  an  APP  calculation  algorithm, in which
the forward-and-backward message passing mechanism of BP  operates  on the subtrellises.
The proposed  BP-MSDD is  capable of significantly  outperforming  the conventional  hard-decision MSDDs.
However,  the    computational   complexity  of the BP-MSDD  increases exponentially with the number of MSDD trellis states.
To circumvent this excessive complexity   for   practical implementations,
we reformulate the   BP-MSDD,
and  additionally propose a Viterbi algorithm (VA)-based hard-decision MSDD (VA-HMSDD) and a  VA-based soft-decision MSDD (VA-SMSDD).
Moreover,   both the proposed BP-MSDD and VA-SMSDD can be exploited in conjunction with  soft  channel decoding to
obtain  powerful   iterative detection and decoding  based  receivers.
Simulation results   demonstrate the effectiveness of the proposed  algorithms   in  DSTBC-aided  UWB-IR systems.

\end{abstract}

% Note that keywords are not normally used for peerreview papers.


\section{Introduction}
\global\long\def\figurename{Fig.}

Coherent detection has been widely investigated and successfully applied in many wireless communication systems  \cite{50years}.
However, coherent detection requires accurate channel state information,  which becomes
expensive or even infeasible  to obtain in   certain scenarios,  such as
 ultra-wideband impulse radio (UWB-IR) systems,
 massive multiple-input multiple-output (MIMO) systems, millimeter communications, dense wireless networks and so on
\cite{50years}-\cite{ energycapture}.
Therefore, noncoherent detection avoiding  channel estimation  has attracted  growing research interests   \cite{ noncoherentdete}, \cite{noncoherent}.
Noncoherent detection  techniques  primarily rely on  differential detection (DD)
 \cite{noncoherent, differential}.
Due to  the doubled variance of the effective noise,  the standard one-symbol   DD  suffers 3dB signal-to-noise ratio (SNR) loss in the
bit-error rate (BER) performance  compared to their   coherent counterparts.
To circumvent this dilemma,  multiple-symbol differential detection (MSDD)
has been advocated as  an effective  algorithm   for  improving the  detection  performance of   differential receivers
\cite{hop}-\cite{tianzhi}.
By exploiting an increased  observation window size, MSDD is capable of flexibly reducing   the   BER performance gap  between noncoherent and coherent detection.

Noncoherent   detection  also leads  to promising low-complexity and energy-efficient receivers  for  multipath fading  channels \cite{efficientcoherent}.
As a milestone for reliable communication over multipath fading channels, multi-antenna  based space-time block-codes (STBC) were invented in \cite{space-time}, which substantially improved the  BER    performance compared to the  single-antenna systems.
Furthermore,  DD-based STBC  (DSTBC)  systems were proposed in  \cite{4v}.
In order to reduce the SNR loss facing the DD scheme of \cite{4v}, an   MSDD   algorithm  was conceived   for DSTBC systems in \cite{5p}.
 Generalized likelihood ratio test based  MSDD  (GLRT-MSDD),
decision feedback  based   MSDD and
sphere decoding  based MSDD (SD-MSDD)
were  investigated    for  the DSTBC aided systems in
\cite{ wtt, block}.
However, the aforementioned MSDD algorithms  are all based on   hard-decision detection,  whose BER performance often remains unsatisfactory.
By comparison, soft-decision  based   iterative MSDD  has been perceived as a more  promising algorithm  for improving the  detection  performance in
channel-coded systems
\cite{iterativenon}-\cite{llr},
where MSDD is  typically concatenated in an iterative manner   \cite{con}  with the decoder of an
error-correcting code (ECC) that is capable of generating soft outputs, such as  convolutional codes,
turbo codes and low-density parity-check codes
  \cite{con}-\cite{yshaoshitcom}.
In iterative MSDD schemes of channel-coded systems, how to calculate the  \emph{ a posteriori} probability (APP)  of each bit transmitted becomes the key problem of interest.

Owing  to the appealing benefit of forward-and-backward message passing,  belief propagation (BP)  has  been successfully applied in many applications, especially in iterative receiver design  \cite{frfacotr2001}-\cite{decoding}.
However, most of them are investigated in the context of coherent detection  systems.
To the best of our knowledge, no BP-based soft-input soft-output (SISO) MSDD has been proposed in the open literature
in  noncoherent DSTBC-aided  UWB-IR  systems.

Against the above  background, in this paper,   our aim is to design a BP-based SISO MSDD framework with its illustrative application in the  DSTBC  aided UWB-IR  systems.  Our particular attention is focused on how to calculate the APPs of the information bits  transmitted.
As such, firstly,  an equivalent channel model  is  proposed   for MSDD  that relies on
sampling with an autocorrelation receiver (AcR) architecture.
Furthermore, we describe the AcR sampling with a trellis structure and then decompose the trellis into multiple subtrellises.
As a beneficial result,  the  bidirectional    message passing of BP is performed  on these   subtrellises,
and a mathematical framework for soft  MSDD, dubbed BP-MSDD, is obtained.
Since  all  subtrellises are taken into account,  our  BP-MSDD  is capable of generating    reliable  \emph{ a posteriori} information.
Notably, the proposed BP-MSDD is capable of achieving more  competitive  detection   performance
than  the existing  GLRT-MSDD     advocated for the DSTBC aided  UWB systems.
However, the  computational  complexity   of   the  BP-MSDD
increases exponentially with the number of MSDD trellis states.
Additionally, as a common feature of  MSDD schemes,
BP-MSDD also exhibits high computational complexity  when  the observation window size increases.
To circumvent the excessive complexity of BP-MSDD,
we further propose a  Viterbi algorithm (VA)-based  hard-decision MSDD (VA-HMSDD)
and a VA-based soft-decision MSDD (VA-SMSDD),
which  constitute attractive practical alternatives  due to their  advantage over BP-MSDD in terms of performance-versus-complexity tradeoff.
Additionally,  the proposed BP-MSDD and VA-SMSDD can be exploited in conjunction with soft  channel decoder   to  conduct   iterative detection and decoding (IDD).
Hence, they  are  particularly attractive for the noncoherent systems.
For example, due to the high frequency selectivity and dense multipaths of UWB-IR channels  \cite{win},
noncoherent UWB-IR systems are preferred because  of their low implementation complexity.
Although the proposed MSDD  algorithms  can generally be used in a number of noncoherent detection scenarios, for the sake of convenience,
in this paper our numerical simulations are carried out in the context of DSTBC aided  UWB-IR systems
to validate the effectiveness of the proposed  algorithms.
For explicit clarity, the main contributions  of this paper  are summarized as follows.

\newcounter{numcount2}
\begin{list}{ \arabic{numcount2})}{\usecounter{numcount2}
\setlength{\itemindent}{0em}\setlength{\rightmargin}{0em}}
\setlength\leftskip{-4ex}

\item
{
An equivalent channel model is developed  with   the  aid of   a rigorous theoretic  analysis   in the  context of
DSTBC  aided  UWB-IR  systems.
This  model  is  attractive, because relying   on this   model,
 accurate soft information  can be    efficiently  obtained  for MSDD.
}

\item
{
Inspired by the forward-and-backward message passing  mechanism of BP,
the  BP-MSDD is conceived,  of which  the message passing  mechanism  is implemented  on  the DSTBC trellis diagram.
In contrast to the existing MSDD algorithms of    \cite{wtt}, \cite{block},
the proposed BP-MSDD exclusively benefits from  its  bidirectional  passing mechanism.
}

\item
{ In order to  circumvent the  excessive complexity  of BP-MSDD for  practical applications,
the survivor paths  of  the   GLRT-MSDD  trellises  are obtained    by  reducing   the number of  trellis states,
and correspondingly,  the  VA-HMSDD is proposed, in which   the maximum tolerable computational   complexity can be restricted
through a  proper selection of the VA memory depth.
This advantage is particularly attractive,  because the  prohibitive computational complexity of the  GLRT-MSDD  trellises
is one of the major challenges   facing the BP-MSDD.

\item
{Upon  extending the trellis-based
detection further,
the  VA-SMSDD is designed for enhancing  the detection performance of VA-HMSDD.
Another benefit of the VA-SMSDD    is that it also enjoys a lower computational
complexity than the   BP-MSDD.
More specifically, the   computational   complexity of BP-MSDD  increases  exponentially  with the number of the trellis states.
By contrast,  VA-SMSDD has a  computational  complexity  increasing  only    linearly with  the number of  trellis states,
although its BER performance is slightly degraded  compared with  that of  BP-MSDD.
}
}\end{list}

We emphasize that the  MSDD algorithms   proposed  in this paper can  be extended and  applied  in
many other communication systems after necessary minor adaptations.
For example, several  MSDD  schemes   have  been developed for the symbol detection in differential phase shift keying (DPSK) aided systems  as well as differential space-time encoded MIMO systems \cite{pisit}. In particular, we consider the  DSTBC aided  UWB-IR system just  as an application for illustrating the principle of the proposed MSDDs. The UWB-IR system exploits  narrow pulses of  nanoseconds to transmit information bits, and the received signal consists of a large number of resolvable multipath components after passing through the UWB-IR channel.
Hence, the stringent requirement  imposed on channel estimation makes it difficult and costly to implement the optimal coherent
receiver. Thus, noncoherent detection becomes inevitable in UWB-IR systems.
In what follows,   the proposed   BP-MSDD, VA-HMSDD  and VA-SMSDD will be  investigated in  the  context of   DSTBC aided  UWB-IR systems.

The rest of this paper is organized as follows.
In Section \uppercase\expandafter{\romannumeral2}, we introduce the   DSTBC system model.
In Section \uppercase\expandafter{\romannumeral3}, we commence with a   brief  review  of the  AcR-based MSDD with the aid of  an equivalent channel model,
then  the  BP-MSDD  is proposed    relying on this  model,
which is followed by formulating both the VA-HMSDD and the  VA-SMSDD for reducing   the  computational   complexity.
Our simulation   results are provided   in Section \uppercase\expandafter{\romannumeral4}.
Finally, conclusions are drawn in Section \uppercase\expandafter{\romannumeral5}.

\noindent\textbf{Notations:}
 Lower-case  (upper-case) boldface symbols denote vectors (matrices);
${\mathbf{I}}_H$ represents  the $H\times H$ identity matrix;
$(\cdot)^\mathrm{T}$  and  $\mathrm{Tr}{(\cdot)}$ represent the transpose  and the trace of a matrix, respectively;
$\| \cdot \|_{\mathrm{F}}$ denotes  the  Frobenius-norm of a matrix;
$\mathrm{E}[\cdot]$ and $\mathrm{Var}[\cdot]$  denote the  expectation and  the  variance of a random variable, respectively;
$*$  stands for convolution;
$ \propto$ means "is proportional to";
$\mathrm{p}(\cdot)$   stands for  the  probability density function   of  a  continuous random variable;
$\mathrm{Pr}(\cdot)$  denotes  a   probability;
 both  $\sum\limits_{\mathbf{U}:\sim \mathbf{U}_{i} } $  and  $\sum\limits_{\sim\mathbf{U}_{i}}$
 represent the   summation of all the  variables    in $\mathbf{U}$  except for   $\mathbf{U}_{i}$.

\section{System Description}

\begin{figure}
\begin{centering}
\includegraphics[width=8cm]{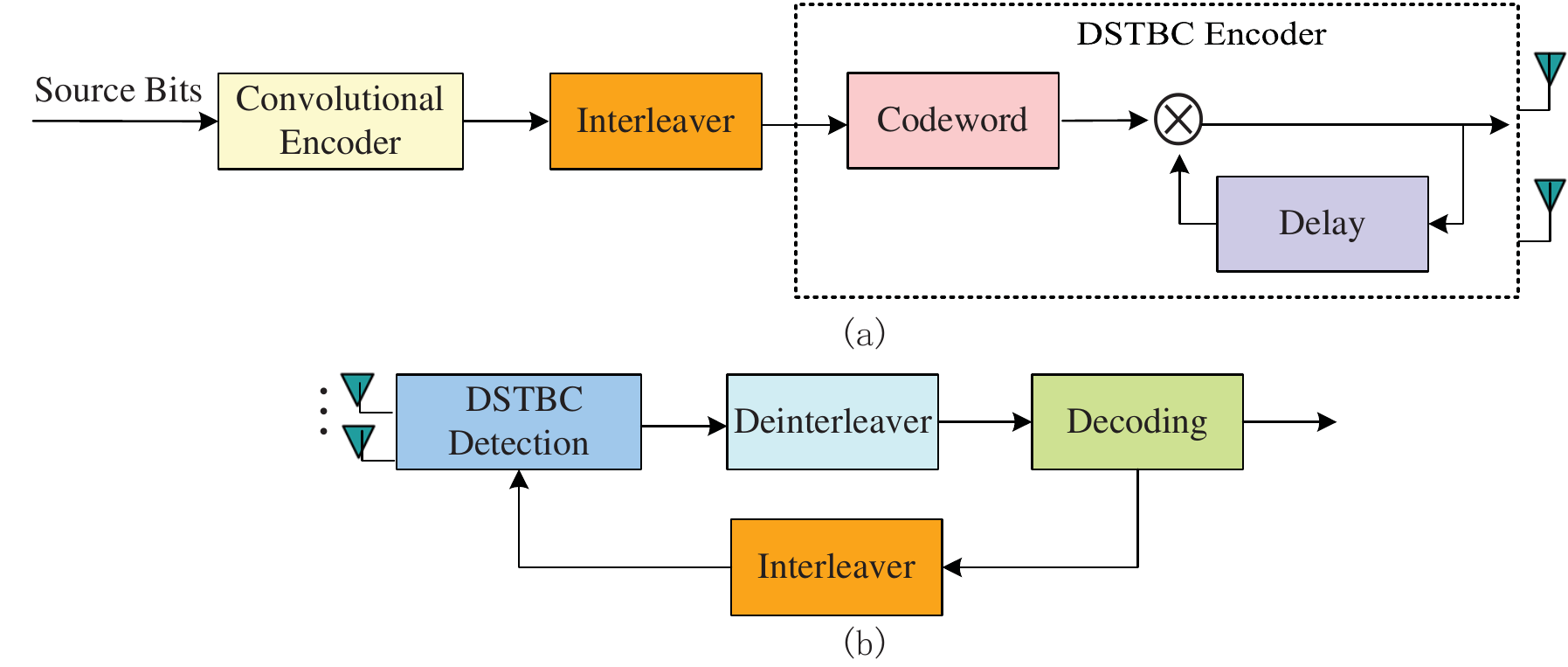}
\caption{\label{system} The diagram of a  DSTBC  system. (a) and (b) denote  a  two-antenna transmitter and a  $Q$-antenna receiver, respectively. }
\par\end{centering}
\end{figure}

For illustrative purpose, we consider a DSTBC aided  UWB-IR  system equipped with  two transmit antennas  and  $Q$ ($ Q\geq1 $) receive  antennas.
The diagram of a basic DSTBC system  is  demonstrated  in   Fig. \ref{system}.
Firstly, the information bits  are encoded with an ECC, e.g.,  a convolutional code.
Then, the coded bits are interleaved and  fed into the DSTBC encoder.
As shown    in \cite{zq},  the channel-encoded  information  bits  are divided into blocks, and then  every two bits
are  mapped onto an
information-bearing space-time coded  symbol.
Each symbol is selected from the code  book $\Omega=\{\mathbf U^{0},\mathbf U^{1},\mathbf U^{2},\mathbf U^{3}\}$ according to the  rule of
$00:\mathbf{U}^{0}\rightarrow
\left(\begin{array}{ccc}
    1 & 0 \\
    0 & 1\\
  \end{array}
\right)$,
$01:\mathbf{U}^{1}\rightarrow \left(
  \begin{array}{ccc}
    -1 & 0 \\
    0 & -1\\
  \end{array}
\right)$,
$10:\mathbf{U}^{2}\rightarrow\left(
  \begin{array}{ccc}
    0& 1 \\
    -1 & 0\\
  \end{array}
\right)$,
$11:\mathbf{U}^{3}\rightarrow\left(
  \begin{array}{ccc}
    0& -1 \\
    1 & 0\\
  \end{array}
\right)$.
Each codeword  $ \mathbf U^{f} \in \Omega, f=0,1,2,3$,  is   constructed according to the property
$\mathbf{U}^{f}(\mathbf{U}^{f})^{\mathrm{T}}=(\mathbf{U}^{f})^{\mathrm{T}}\mathbf{U}^{f}=\mathbf{I}_{2}$.
Note that  each   codeword employed in the DSTBC system  has to be  a  unitary matrix.
It is well known that for a unitary matrix,
each row (column) and the other rows (columns)  are mutually orthogonal.
When the unitary matrices  are designed according to the DSTBC scheme of   \cite{space-time},
the  proposed  algorithms  can be extended to  systems  having   more than two transmit antennas.
Furthermore, upon invoking  the differential encoding,
the  coded  symbol to be transmitted   is obtained by
\begin{equation}\label{transmittedd}
\mathbf{D}_{i+1}=\mathbf{D}_{i}\mathbf{U}_{i+1},
\end{equation}
where the space-time coded  information  symbol satisfies  $\mathbf{U}_{i+1}\in\Omega$,
the $2\times2$ matrix $\mathbf{D}_{i+1}$  is  sent with   two antennas during two frame durations,  the reference symbol for differential encoding  is set to
$\mathbf{D}_{0}=\left(
  \begin{array}{cc}
   1 & -1\\
   1& 1 \\
  \end{array}
\right)$,
$i=0, 1, \cdots, N-1$,  and $N$ represents the number of differentially encoded symbols to be transmitted.
Let $d_{p,2i+n-1}$  denote  the entry in  the  $p$th  row and the
 $n$th   column of the  matrix $\mathbf{D}_{i}$,  where $p=1, 2$  and  $n=1,2$.
Then,   the  signal   transmitted   from the $p$th antenna  is given by
\begin{equation}
\begin{split}
s_{p}(t)&=\sum\limits_{i=0}^{ N-1}\sum\limits_{n=1}^{2}d_{p,2i+n-1}\omega(t-(n-1)T_{\mathrm{f}}-iT_{\mathrm{s}})\\
&=\sum\limits_{i=0}^{N-1}\sum\limits_{n=1}^{2}d_{p,2i+n-1}\omega(t-(2i+n-1)T_{\mathrm{f}})\\
&=\sum\limits_{j=0}^{2N-1}d_{p,j}\omega(t-jT_{\mathrm{f}}),
\end{split}
\end{equation}
where $\omega(t)$ denotes the monocycle pulse   that   is real-valued waveform  with a very short duration
$T_{\mathrm{\omega}}$; $t$ is the continuous time index;
$T_{\mathrm{f}}$
 is the frame duration, which is usually hundreds to thousands times of $T_{\mathrm{\omega}}$ \cite{tianzhi};
a single transmitted-symbol duration is  $T_{\mathrm{s}}=2T_{\mathrm{f}}$,
which implies that
in the UWB-IR system considered,   two  frames are employed  to transmit one  space-time symbol, and each frame includes one very short pulse;
the   discrete
 time index $j=2i+n-1$  is introduced to replace the
 two-dimensional
   index $(i,n)$ and  $d_{p,2i+n-1}$ is rewritten as $d_{p,j}$, correspondingly.
We consider  a quasi-static dense multipaths fading environment \cite{win}, then the  UWB channel impulse response between the
  $p$th
  transmit antenna and the $q$th
($1\leq q\leq Q$) receive antenna  is  formulated  as
\begin{equation}
h_{p,q}(t)=\sum\limits_{l=1}^{L_{p,q}}\alpha_{l}^{p,q}\delta(t-\tau_{l}^{p,q}),
\end{equation}
where  $L_{p,q}$ denotes the number of propagation paths;  $\alpha_{l}^{p,q}$  and $\tau_{l}^{p,q}$ are  the path-gain coefficient  and the delay associated with the
$l$th  path, respectively; and  $\delta(t)$  is  Dirac delta  function.
As a result,   the overall channel response   between the
$p$th  transmit antenna and the
  $q$th  receive antenna is expressed as
\begin{equation}
g_{p,q}(t)=\omega(t) * h_{p,q}(t)=\sum\limits_{l=1}^{L_{p,q}}\alpha_{l}^{p,q}\omega(t-\tau_{l}^{p,q}).
\end{equation}
Then, the signal at the  $q$th
 receive antenna is  given   by
\begin{equation}\label{yt}
\begin{split}
y_{q}(t)&=\sum\limits_{p=1}^{2}s_{p}(t) * h_{p,q}(t)+n_{q}(t)\\
&=\sum\limits_{p=1}^{2}\sum\limits_{j=0}^{2N-1}d_{p,j}g_{p,q}\left(t-
jT_{\mathrm{f}}\right)+n_{q}(t),
\end{split}
\end{equation}
where  $n_{q}(t)$ denotes  the additive white Gaussian noise (AWGN)   having   zero mean  and two-sided power spectral density $N_{0}/2$.

\section{  BP-MSDD for  DSTBC  Aided  UWB-IR Systems}

Several  noncoherent detection  schemes   have  been developed in the context of    UWB  systems in  \cite{wtt}, \cite{block},  \cite{zq}.
Note that  most  of these MSDD schemes are based on hard-decision, which
results  in   unsatisfactory   detection performance   in many applications.
In general, soft-decision  based MSDDs are expected to be capable of outperforming their
hard-decision  based counterparts.
However, soft MSDD remains an inadequately studied topic and few contributions are found in open literature
\cite{pisit}-\cite{ llr}.
A linear prediction algorithm   was   proposed in \cite{pisit} to utilize the temporal correlation of fading
 and  a soft information aided iterative multiple-symbol  noncoherent detection scheme was conceived for differential unitary space-time modulation in Rayleigh flat fading channels.
Furthermore,  the authors of   \cite{mars}   developed
an  iterative  detector for  coded differential phase shift keying  (DPSK) modulated signals in  an AWGN channel with unknown phase.
Additionally,  a  simplified  iterative SD-MSDD  algorithm  was  developed   for DPSK  modulated signals
in \cite{llr}, where a Rayleigh-fading channel was considered.
These existing soft MSDDs  can  provide  soft  information   with acceptable quality   for detection.
However,   it is necessary to develop  a  more sophisticated   soft MSDD   algorithm,
which ought to be    capable of achieving  much better   detection performance.
In this section,   a soft-decision  based
 BP-MSDD is proposed to tackle this problem  in  the context of  DSTBC aided  UWB-IR systems.
Firstly,  an equivalent channel model is presented  for  MSDD  relying on a novel  AcR structure, where the sampling mechanism   is judiciously   modified
and the observation window slides only  a single
symbol duration  $T_{\mathrm{s}}$
 each time.
This sampling mechanism is different from  the $M$-symbol-duration sliding  sampling mechanism   used    in \cite{pisit}.
Then, BP-MSDD is proposed by exploiting this equivalent
model and the forward-and-backward message passing mechanism.

\subsection{ Equivalent Channel Model }

\begin{figure}
\begin{centering}
\includegraphics[width=8cm]{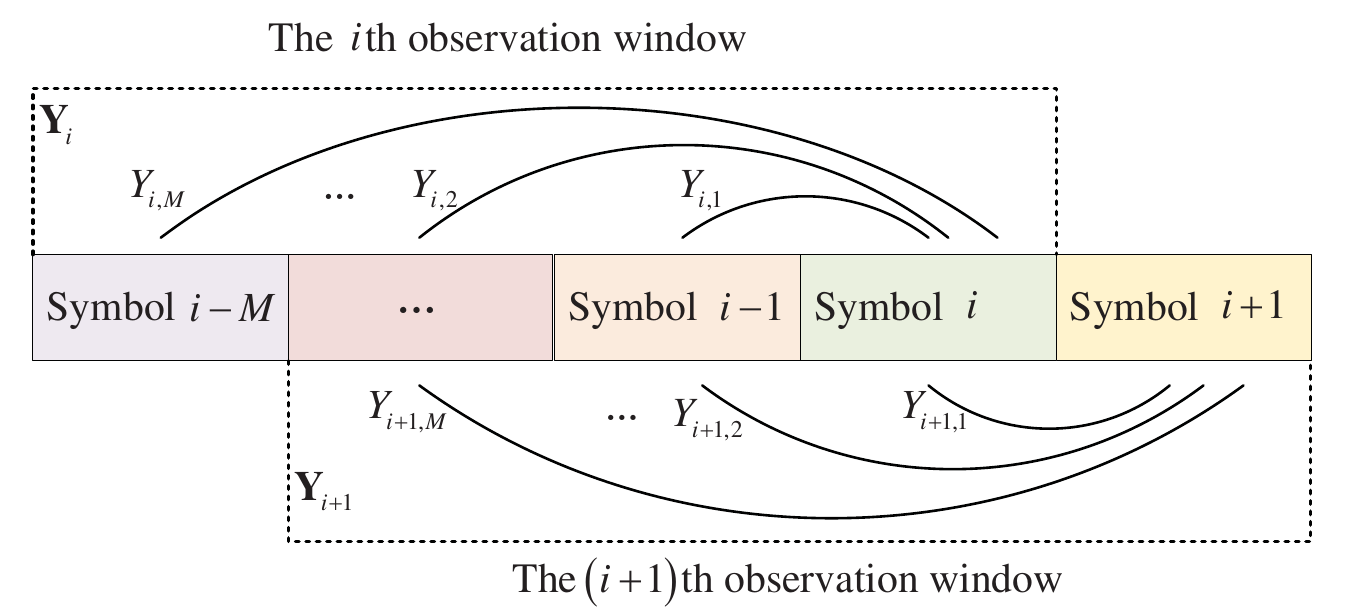}
\caption{\label{window} An illustration of  the proposed AcR sampling mechanism for MSDD   in the $i$th  and
$(i+1)$th  observation window. }
\par\end{centering}
\end{figure}

Let us commence with  an AcR  architecture that  is developed for MSDD sampling.
For  convenience of exposition, the observation window size is set to $M+1$ $ (M \geq1) $ symbol durations, during which  the channel is assumed to  remain invariant
\cite{block},  \cite{zq}.
In  \cite{block},  the observation window  slides forward  $M$ symbol durations  when the current $M$ symbols have been detected.
This is indeed a  block-by-block sampling scheme,  where  the correlations amongst  the symbols of   different blocks are unfavorably ignored.
Motivated by this observation,   we obtain a modified symbol-by-symbol sliding mode,  in which
the observation window slides   forward  one symbol duration
$T_{\mathrm{s}}$  each time.
This symbol-by-symbol   sampling   scheme is particularly suitable for   exploiting  the correlations  among
symbols of different  blocks, since  the detection of  each   symbol benefits from
capitalizing on    the  information of   all the other  symbols.
In  Fig. \ref{window},  we   illustrate  the   details of the  proposed sampling scheme  by examining   two adjacent observation windows.
To gain more in-depth insights,   we formulate the proposed AcR sampling mechanism in a rigorous mathematical manner,
and then elaborate on its  significance for  BP-MSDD.
Specifically, in  the   $i$th
observation window represented by   $(i-M)T_{\mathrm{s}}\leq t \leq iT_{\mathrm{s}}$,
the samples of the received signal are given by
$\mathbf{Y}_{i}=[\mathbf{Y}_{i,1},\mathbf{Y}_{i,2},\ldots, \mathbf{Y}_{i,M}]$,
where  $\mathbf{Y}_{i,m}$ $(m=1,  \ldots , M)$
is the correlation function   of the received signal and it is  formulated as
\begin{equation}\label{y5}
\mathbf{Y}_{i,m}=\sum\limits_{q=1}^{Q}\int_0 ^ { T_{\mathrm{i}}} \left(\mathbf{y}_{i}^{q}(t)\right)^{\mathrm{T}}\mathbf{y}_{i-m}^{q}(t)dt,
\end{equation}
where $T_{\mathrm{i}}\leq T_{\mathrm{s}}$
    is  the  integration interval,
$ \mathbf{y}_{i}^{q}(t)=[y_{q}(t+2i T_{\mathrm{f}}) \quad y_{q}(t+(2i+1) T_{\mathrm{f}})]$ is  the  signal  waveform vector received   at
 $q$th  antenna.
The  $(u,v)$th
entry   of $\mathbf{Y}_{i,m}$ can be expressed as
$Y_{i,m}(u,v) = \sum\limits_{q=1}^{Q}\int_0 ^ {T_{\mathrm{i}}}y_{q}  [t+2i T_{\mathrm{f}}+(u-1)T_{\mathrm{f}}] y_{q}[t+2(i-m) T_{\mathrm{f}}+(v-1)T_{\mathrm{f}}]dt$
with  $u,v=1,2$.
Substituting (\ref{yt}) into $\mathbf{Y}_{i,m}$, we obtain
\begin{equation}
\begin{split}
Y_{i,m}(u,v)&= \sum\limits_{q=1}^{Q}\int_0 ^ { T_{\mathrm{i}}} \left(\sum\limits_{p=1}^{2}d_{p,2i+u-1}g_{p,q}(t)+n_{i,u}^{q}(t)\right)\\
&\times \left(\sum\limits_{p=1}^{2}d_{p,2(i-m)+v-1}g_{p,q}(t)+n_{(i-m),v}^{q}(t)\right) dt,
\end{split}
\end{equation}
where  $n_{i,u}^{q}(t)=n_{q} [t+(2i+u-1) T_{\mathrm{f}}]$,
and  $n_{(i-m),v}^{q}(t)=n_{q} [t+(2i-2m+v-1)T_{\mathrm{f}}]$
denote the time-shifted noises.
\begin{figure}
\begin{centering}
\includegraphics[width= 7 cm]{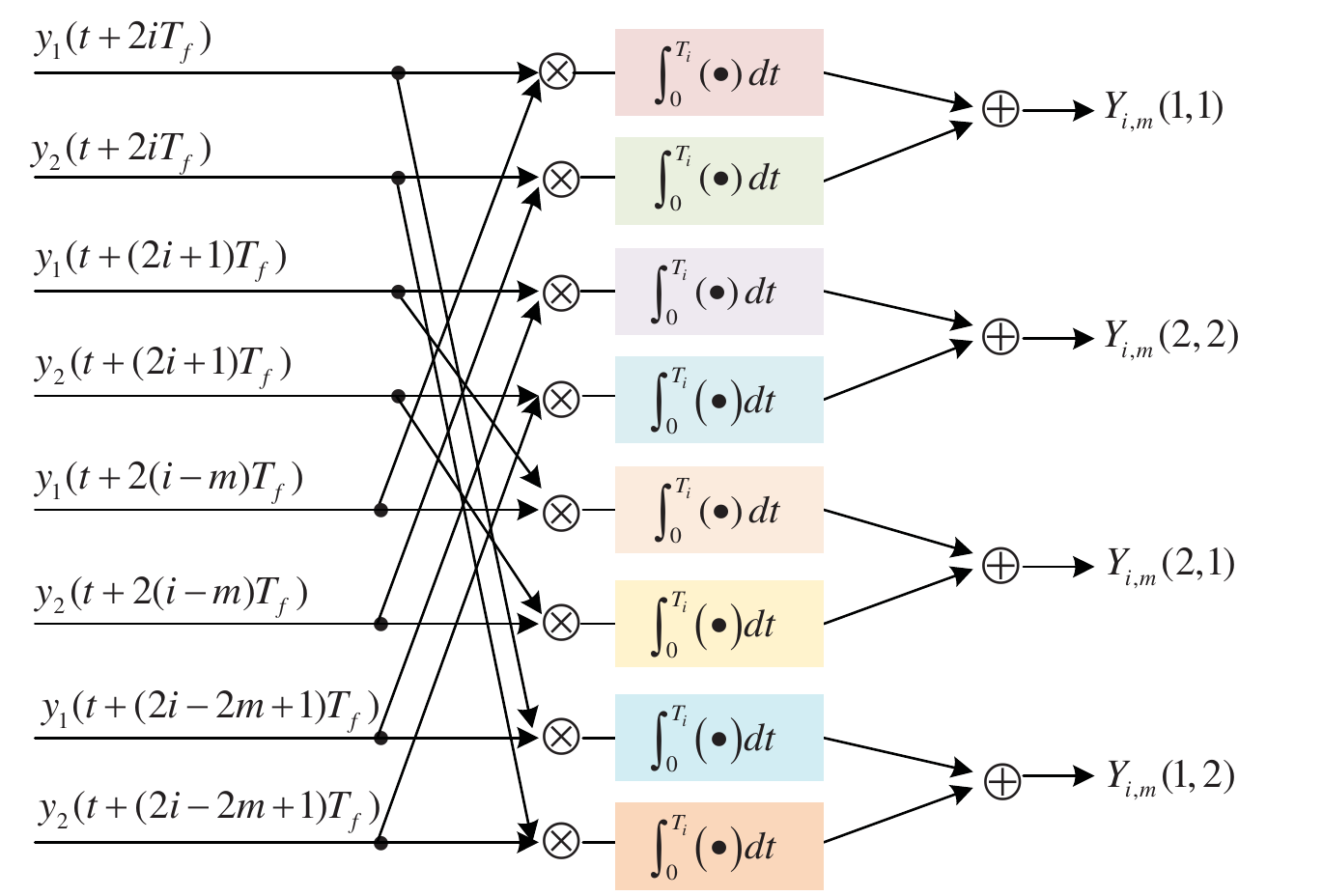}
\caption{ \label{receiver}
The  illustration of the AcR structure  for the $i$th  and the  $(i-m)$th  received symbol matrices  of   DSTBC aided  UWB-IR system  having   two receive antennas.}\label{B}
\par\end{centering}
\end{figure}
The calculation of the   correlation  matrix  (\ref{y5})  requires operations between different segments of the received symbols.
Consider   $Q=2$ as an example, in Fig. \ref{receiver}  we  illustrate  the AcR   structure for the  $i$th
and the
$(i-m)$th
 received  symbol matrices
of    DSTBC aided   UWB-IR system.
For the sake of  convenience,  we divide ${Y}_{i,m}(u,v)$ into  the signal component $S_{i,m}(u,v)$ and  the noise component $N_{i,m}(u,v)$.
Then,   ${Y}_{i,m}(u,v)$ is  reformulated as
\begin{equation}\label{10}
Y_{i,m}(u,v)=S_{i,m}(u,v)+ N_{i,m}(u,v),
\end{equation}
where  $S_{i,m}(u,v)$  is given by
\begin{equation}
\begin{split}
&S_{i,m}(u,v)= \\
& \sum\limits_{q=1}^{Q}\int_{0}^{ T_{\mathrm{i}}
}\Bigg(\sum\limits_{p=1}^{2}d_{p,2i+u-1}g_{p,q}(t)
\sum\limits_{p=1}^{2}d_{p,2(i-m)+v-1}g_{p,q}(t)\Bigg)dt,
\end{split}
\end{equation}
and  $N_{i,m}(u,v)$  can be  expressed as
\begin{equation}
N_{i,m}(u,v)= N_{1}+N_{2}+N_{3},
\end{equation}
with
\begin{equation}\label{n1}
N_{1}=\sum\limits_{p=1}^{2}d_{p,2i+u-1}\sum\limits_{q=1}^{Q}\int_{0}^{ T_{\mathrm{i}}}g_{p,q}(t)n_{(i-m),v}^{q}(t)dt,
\end{equation}
\begin{equation}\label{n2}
N_{2}=\sum\limits_{p=1}^{2}d_{p,2(i-m)+v-1}\sum\limits_{q=1}^{Q}\int_{0}^{T_{\mathrm{i}}}g_{p,q}(t)n_{i,u}^{q}(t)dt,
\end{equation}
and
\begin{equation}\label{n3}
N_{3}=\sum\limits_{q=1}^{Q}\int_{0}^{ T_{\mathrm{i}}}n_{i,u}^{q}(t)n_{(i-m),v}^{q}(t)dt.
\end{equation}
Similar to that of \cite{zq}, we assume  that $n_{q}(t)$ is
an  AWGN  process with bandwidth  $W \gg 1/T_{\mathrm{i}}$.
Clearly, $N_{1}$ and $N_{2}$ are Gaussian random variables, while   $N_{3}$
 approximately obeys a Gaussian distribution \cite{wtt}.
Given the channel realizations  $h=\{h_{p,q}(t)\}$,
according to the central limit theorem,  $N_{i,m}(u,v)$ in (\ref{10}) can be considered as a noise which has    zero mean  and conditional variance
\begin{equation}\label{sigma2}
\sigma^{2}=\mathrm{Var}[N_{i,m}(u,v)|h]= \mathrm{E}[N_{1}^{2}|h]+\mathrm{E}[N_{2}^{2}|h]+\mathrm{E}[N_{3}^{2}].
\end{equation}
For analytical convenience,  we define  $\varepsilon_{p,q}=\int_{0}^{T_{\mathrm{i}}}h_{pq}^{2}(t)dt$, and
 $\xi=\sum\limits_{p=1}^{2}\sum\limits_{q=1}^{Q}\varepsilon_{p,q}$.
According to (\ref{n1}), (\ref{n2})  and (\ref{n3}), the conditional variance of $N_{1}$, $N_{2}$ and $N_{3}$ can be  expressed   as
$\mathrm{E}[N_{1}^{2}|h]=\mathrm{E}[N_{2}^{2}|h]=\frac{N_{0}}{2}\sum\limits_{p=1}^{2}\sum\limits_{q=1}^{Q}\varepsilon_{p,q}$
and
$\mathrm{E}[N_{3}^{2}]=\frac{QW T_{\mathrm{i}} N_{0}^{2}}{2}$,
respectively.
Substituting  $\mathrm{E}[N_{1}^{2}|h]$,  $ \mathrm{E}[N_{2}^{2}|h]$ and $\mathrm{E}[N_{3}^{2}]$  into (\ref{sigma2}),  we obtain
\begin{equation}\label{sigmazuizhong}
\sigma^{2}= N_{0}\sum\limits_{p=1}^{2}\sum\limits_{q=1}^{Q}\varepsilon_{p,q}+\frac{QWT_{\mathrm{i}}N_{0}^{2}}{2}=N_{0}\xi+\frac{QWT_{\mathrm{i}}N_{0}^{2}}{2}.
\end{equation}
Consequently,  $\mathbf{Y}_{i,m}$  in (\ref{y5}) can be rewritten as
\begin{equation}\label{myim}
\mathbf{Y}_{i,m}=\mathbf{S}_{i,m}+\mathbf{N}_{i,m},
\end{equation}
where
$\mathbf{S}_{i,m}=\left(
  \begin{array}{cc}
   S_{i,m}(1,1) & S_{i,m}(1,2)\\
   S_{i,m}(2,1)& S_{i,m}(2,2)\\
  \end{array}\right)$
and
$\mathbf{N}_{i,m}=\left(
  \begin{array}{cc}
   N_{i,m}(1,1) & N_{i,m}(1,2)\\
   N_{i,m}(2,1)& N_{i,m}(2,2)\\
  \end{array}\right)$
represent  the  signal component  and the   noise component, respectively.
Moreover,  based on the  differential modulation  $\mathbf{D}_{i}=\mathbf{D}_{i-1}\mathbf{U}_{i}$  and $\mathbf{D}_{i}\mathbf{D}_{i}^{\mathrm{T}}=2 \mathbf{I}_{2}$,
we have
\begin{equation}
\begin{split}\label{differ}
&\mathbf{D}_{i}\mathbf{D}_{i-m}^{\mathrm{T}}=\mathbf{D}_{i-1}\mathbf{U}_{i}\mathbf{D}_{i-m}^{\mathrm{T}}  \\
&=\mathbf{D}_{i-m}\mathbf{U}_{i-m+1}\cdots \mathbf{U}_{i-1}\mathbf{U}_{i}\mathbf{D}_{i-m}^{\mathrm{T}}=2\prod_{z=i-m+1}^{i}\mathbf{U}_{z}.
\end{split}
\end{equation}
Jointly considering  (\ref{myim}) and (\ref{differ}), we can concisely reformulate   $\mathbf{Y}_{i,m}$   as
\begin{equation}\label{yim21}
\mathbf{Y}_{i,m}=(\mathbf{D}_{i}\mathbf{D}_{i-m}^{\mathrm{T}}) \xi+ \mathbf{N}_{i,m}=(2\prod_{z=i-m+1}^{i}\mathbf{U}_{z})\xi+ \mathbf{N}_{i,m}.
\end{equation}
Therefore,  the conditional probability required by the  MSDD of DSTBC aided UWB-IR systems
in a single  observation window is   formulated as
\begin{equation}\label{cond}
\begin{split}
&\mathrm{p}(\mathbf{Y}_{i}|\mathbf{U}_{i-M+1},  \cdots, \mathbf{U}_{i-1}, \mathbf{U}_{i})\propto  \\
& \prod_{m=1}^{M}\mathrm{exp}\left(-\frac{\| \mathbf{Y}_{i,m}- (2\prod_{z=i-m+1}^{i}\mathbf{U}_{z})\xi \|_{\mathrm{F}}^{2}}{\sigma^{2}}\right),
\end{split}
\end{equation}
which will be applied into generating    reliable  \emph{ a posteriori} information for the    BP-MSDD  with soft-decisions.
We assume that  $T_{\mathrm{s}}= N_{\mathrm{f}}\cdot T_{\mathrm{f}}$,
then
 in   (\ref{cond}) the receivers can employ the energy  parameter  as
$\hat{\xi}=\sqrt{\sum_{i,m}\mathbf{Y}^{2}_{i,m}}/N_{\mathrm{f}}$ \cite{jsac},
 where $N_{\mathrm{f}}$ is the number of frames in a single symbol duration  $T_{\mathrm{s}}$.
After obtaining    the sampling   result   $\mathbf{Y}_{i}$,
the next sampling outcome  $\mathbf{Y}_{i+1}$  may be   calculated  based on   the next observation window
$(i-M+1)T_{\mathrm{s}}\leq t \leq (i+1)T_{\mathrm{s}}$.

\noindent\textbf{Remark 1:}
The conditional probability (\ref{cond}) can be regarded as a metric function.
Relying on the proposed sampling mechanism and metric function of (\ref{cond}), our  BP-MSDD is derived  in the next subsection.
The key  insight behind the derivation of the BP-MSDD is that we have to construct
 a  graphical   model for the whole   sampling packet,
where our beliefs in  the information  symbols  propagate  throughout the  graph with  the aid of the  BP message passing mechanism.

\begin{figure}
\begin{centering}
\includegraphics[width= 5 cm]{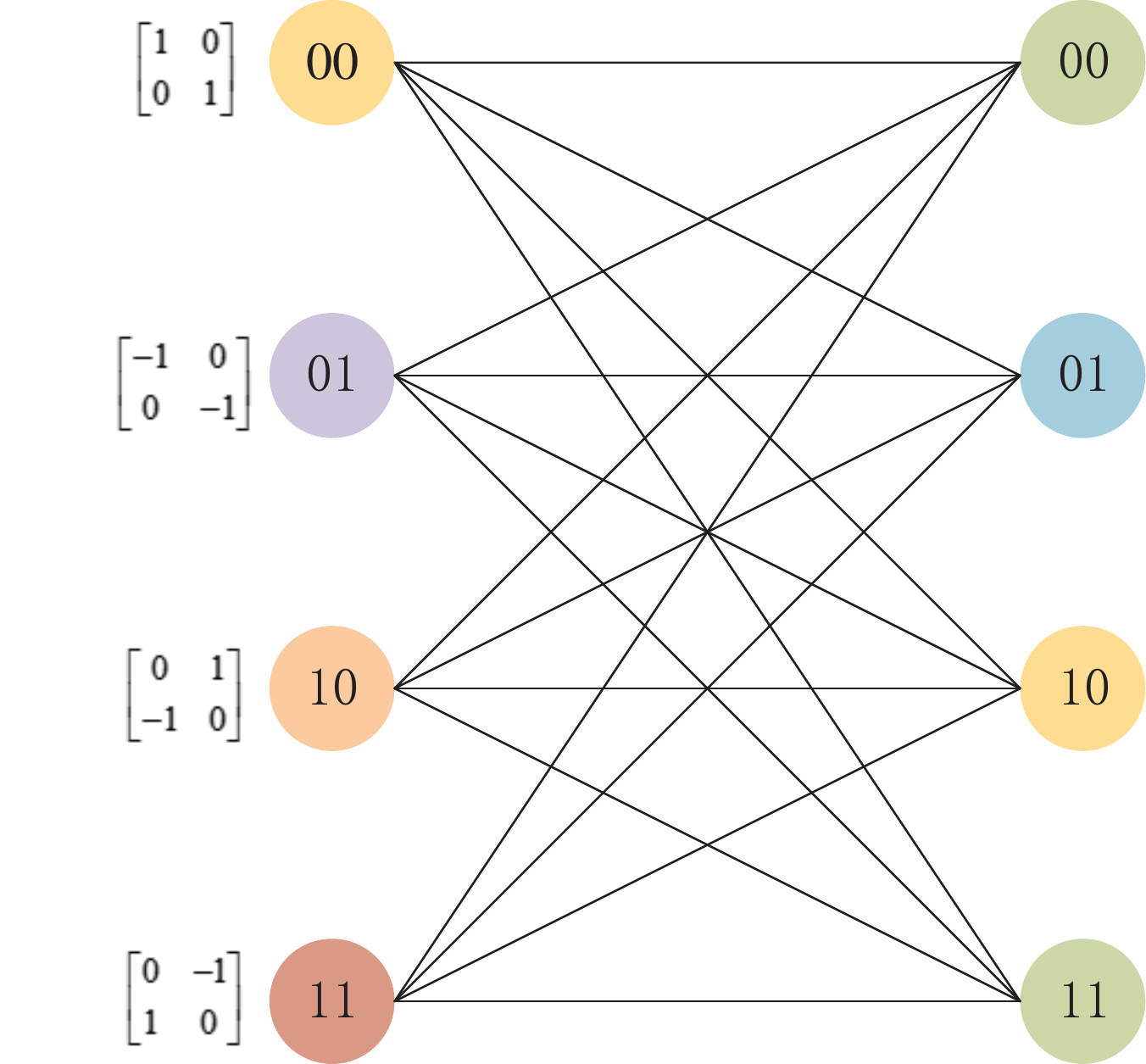}
\caption{ \label{DSTBCtrellis}
The  DSTBC trellis diagram, where the nodes represented by circles denote the states
and the lines represent the  branches corresponding to  the state transitions.
The matrices on the left represent the input information symbols, and the state transition is from the left to the right.
}\label{B}
\par\end{centering}
\end{figure}

\subsection {Belief Propagation Based MSDD}

Aiming for   obtaining the information bits' maximum APPs,  i.e., we use the  maximum \textit{a posteriori} probability (MAP) criterion,
our  BP-MSDD  focuses  on  the calculations of  the  bitwise APPs      that are denoted by
$\Lambda[d_{i}(k)]=\mathrm{p}[d_{i}(k)=z|\mathbf{Y}]$,
where $d_{i}(k)$ is the $k$th  bit of  the  information symbol  $\mathbf{U}_{i}$,    $k=1,2$ and  $z=0,1$.
Let   $\mathbf{Y} \triangleq  \{\mathbf{Y}_{i}\} $
be the matrix  containing the  sampling results of all the observation windows
and  $\mathbf{U} \triangleq \{\mathbf{U}_{1},  \mathbf{U}_{2}, \cdots \mathbf{U}_{N}\}  $  containing all the information  symbols.
Then,   the   APP  for   $\mathbf{U}_{i}$  is given by
\begin{equation}\label{ap}
\mathrm{Pr}(\mathbf{U}_{i}|\mathbf{Y})=\frac{\mathrm{Pr}(\mathbf{U}_{i},\mathbf{Y})}{\mathrm{p}(\mathbf{Y})}
=\frac{\mathrm{p}(\mathbf{Y}|\mathbf{U}_{i})\mathrm{Pr}(\mathbf{U}_{i})} {{\mathrm{p}(\mathbf{Y})}}.
\end{equation}
Thus, $\mathbf{U}_{i}$  can be   determined with the aid of the MAP criterion as
\begin{equation}\label{aparg}
\begin{split}
&(\mathbf{U}_{i})_{\mathrm{MAP}}=\mathrm{arg} \max \limits_{\mathbf{U}_{i}} \mathrm{Pr}(\mathbf{U}_{i}|\mathbf{Y})  \\
&=\mathrm{arg} \max \limits_{\mathbf{U}_{i}}\frac{\mathrm{p}(\mathbf{Y}|\mathbf{U}_{i})\mathrm{Pr}(\mathbf{U}_{i})} {{\mathrm{p}(\mathbf{Y})}}
=\mathrm{arg} \max \limits_{\mathbf{U}_{i}}{\mathrm{p}(\mathbf{Y}|\mathbf{U}_{i})\mathrm{Pr}(\mathbf{U}_{i})},
\end{split}
\end{equation}
where $\mathrm{p}(\mathbf{Y})$  can be   removed,  because it is independent of   $\mathbf{U}_{i}$.
According to (\ref{aparg}) and  the relationship between the  joint probability density function and the marginal probability density function  [40], we can   obtain
\begin{equation}\label{app}
\mathrm{Pr}(\mathbf{U}_{i}|\mathbf{Y})\propto\sum_{\mathbf{U}:\sim \mathbf{U}_{i}} \mathrm{p}(\mathbf{Y}|\mathbf{U})\mathrm{Pr}(\mathbf{U}),
\end{equation}
which means
$\mathrm{Pr}(\mathbf{U}_{i}|\mathbf{Y})\propto  \sum_{\mathbf{U}_{1}, \mathbf{U}_{2}, \cdots, \mathbf{U}_{i-1}, \mathbf{U}_{i+1}, \cdots, \mathbf{U}_{N}} \mathrm{p}(\mathbf{Y}|\mathbf{U})\mathrm{Pr}(\mathbf{U})$  for any
$\mathbf{U}_{i}\in \mathbf{U}$.
The direct  calculation of (\ref{app})  imposes a complexity increasing exponentially with  $N$.
Hence, it is challenging to directly implement it in  practical applications when  $N$  is large.
A reduced-complexity   strategy   relying  on   trellis factorization  was  proposed  to address this  problem   \cite{memory}.
To elaborate a little further, it was shown in \cite{memory} that the global probability function can be factorized into a number of local functions to facilitate the mitigation of inter-symbol interference.
Inspired  by this  innovative idea,
we  develop  a sophisticated   framework  that is   dedicated to  noncoherent detection.
More specifically, the calculation of  (\ref{app}) can be reformulated  as a problem of
estimating the   APPs   of the state  transitions   of   a Markov source \cite{1order},
which  constitutes   a discrete  finite-state Markov process  operating over   a noisy discrete memoryless channel (DMC).
The indices of the  distinct states  are indexed by the integer  $c$.
At the  time instant $t'$, the source state is denoted by $\mathbf{S}_{t'}$
and its output is  ${X}_{t'}$.
A  sequence generated by  the source  is obtained from the  state transitions, and
the  state transition probability  of   $\mathbf{S}_{t'-1}\rightarrow \mathbf{S}_{t'}$
  is  described by
$\mathrm{Pr}(c|c')=\mathrm{Pr}(\mathbf{S}_{t'}=c|\mathbf{S}_{t'-1}=c')$
and  the probability of  its output is given by
\begin{equation}\label{xmm'}
\mathrm{Pr}(X|c',c)=\mathrm{Pr}(X=X_{t'}|\mathbf{S}_{t'-1}=c',\mathbf{S}_{t'}=c).
\end{equation}
In Eq. (24), the values of the states, represented by c' and c, are certain elements of the state space of the Markov chain. Between each two distinct states, there is a transition probability, which can take any arbitrary value in the interval of [0,1]. For the  Markov source,
when  $X_{1}^{\tau}=[X_{1}, X_{2}, \cdots, X_{\tau}]$
is  the  input to  a noisy DMC,  the output is  $Y_{1}^{\tau}=[Y_{1}, Y_{2}, \cdots, Y_{\tau}]$.
For  $ 1 \leq t' \leq \tau$,
the transition probability  can be derived as   \cite{1974}
\begin{equation}\label{dmc}
\mathrm{Pr}(Y_{1}^{t'}|X_{1}^{t'})= \prod_{\eta=1}^{t'}\mathrm{Pr}(Y_{\eta}|X_{\eta}).
\end{equation}
Based on  the above insights,   in the following  an efficient   algorithm  is  developed  to factorize the complicated global function  of  (\ref{app}).
This factorization is  useful for constructing the  trellis representation of MSDD,
as  shown in    Fig. \ref{DSTBCtrellis}.
In order to obtain the    trellis factorization \cite{hidden}  corresponding to   MSDD,
we  define the trellis as follows.
Assuming that  the  state tuple is $\mathbf{S}_{i-1}\triangleq [\mathbf{U}_{i-M}, \cdots, \mathbf{U}_{i-1}]^{\mathrm{T}}$,
if the input symbol is $\mathbf{U}_{i}$,  the output becomes
$\mathbf{X}_{i}\triangleq [ \mathbf{U}_{i-M+1},  \cdots, \mathbf{U}_{i-1}, \mathbf{U}_{i} ]^{\mathrm{T}}$.
The   $i$th
 sampling  $ \{\mathbf{Y}_{i}\} $
 is related to   information  symbols    $\{\mathbf{U}_{i-m} | m=0, \cdots, M-1 \}$.
In the context of   the   discrete  channel, $\mathrm{p}(\mathbf{Y}| \mathbf{U})$ can be factorized as
\begin{equation}\label{puy}
\mathrm{p}(\mathbf{Y}|\mathbf{U})=\prod_{i=1}^{N}\mathrm{p}(\mathbf{Y}_{i}|\mathbf{U}_{i-M},  \cdots, \mathbf{U}_{i-1}, \mathbf{U}_{i}),
\end{equation}
where  $\mathrm{p}(\mathbf{Y}_{i}|\mathbf{U}_{i-M},  \cdots, \mathbf{U}_{i-1}, \mathbf{U}_{i}) $
is given by  (\ref{cond}).
Correspondingly,
$\mathrm{Pr}(\mathbf{U})$  is  reformulated as
\begin{equation}\label{fact2}
\begin{split}
&\mathrm{Pr}(\mathbf{U})=\\
& \prod_{i=1}^{N}\mathrm{Pr}(\mathbf{U}_{i-M},  \cdots, \mathbf{U}_{i-1}, \mathbf{U}_{i}|\mathbf{U}_{i-M-1},  \cdots, \mathbf{U}_{i-2}, \mathbf{U}_{i-1}).
\end{split}
\end{equation}
Furthermore, the  local check function for the state   transition    $\mathbf{S}_{i-1}\rightarrow \mathbf{S}_{i}$   is given by
\begin{equation} \label{trans}
\mathrm{T}_{i}(\mathbf{U}_{i}, \mathbf{X}_{i}, \mathbf{S}_{i-1},\mathbf{S}_{i})
=\left\{
\begin{array}{ll}
1, \quad \mathrm{if } \quad \mathrm{this} \quad \mathrm{event} \quad \mathrm{exists} \\
0,  \quad \mathrm{otherwise},
\end{array}\right.
\end{equation}
which is also proportional to the    state  transition  probability. Explicitly, we have
\begin{equation}
\begin{split}
\label{statetrans}
&\mathrm{T}_{i}(\mathbf{U}_{i}, \mathbf{X}_{i}, \mathbf{S}_{i-1},\mathbf{S}_{i}) \propto \\
&\mathrm{Pr}(\mathbf{U}_{i-M},  \cdots, \mathbf{U}_{i-1}, \mathbf{U}_{i}|\mathbf{U}_{i-M-1},  \cdots, \mathbf{U}_{i-2}, \mathbf{U}_{i-1}).
\end{split}\end{equation}
Based on  (\ref{fact2}), (\ref{trans})  and (\ref{statetrans}),   $\mathrm{Pr}(\mathbf{U})$  satisfies
\begin{equation}\label{pu}
\mathrm{Pr}(\mathbf{U})\propto \prod_{i=1}^{N} \mathrm{T}_{i}(\mathbf{U}_{i}, \mathbf{X}_{i}, \mathbf{S}_{i-1},\mathbf{S}_{i}).
\end{equation}
Defining  $ \mathrm{Pr}(\mathbf{U}_{i})$   as  the \emph{a priori}  probability  of   $\mathbf{U}_{i}$
and  substituting (\ref{puy}),  (\ref{pu}) into (\ref{app}),
we have
\begin{equation}
\label{zuipai}
\begin{split}
&\mathrm{Pr}(\mathbf{U}_{i}|\mathbf{Y})   \varpropto  \\
& \underbrace{\left( \sum\limits_{\mathbf{U}_{1:i-1}} \prod_{w=1}^{i-1} \mathrm{Pr}(\mathbf{U}_{w})
\mathrm{p}(\mathbf{Y}_{w}|\mathbf{X}_{w}) \mathrm{T}_{w}(\mathbf{U}_{w}, \mathbf{X}_{w}, \mathbf{S}_{w-1}, \mathbf{S}_{w})\right)  }_{\triangleq \alpha (\mathbf{S}_{i-1})}\times\\
& \mathrm{Pr}(\mathbf{U}_{i})\mathrm{p}(\mathbf{Y}_{i}|\mathbf{X}_{i}) \mathrm{T}_{i}(\mathbf{U}_{i}, \mathbf{X}_{i}, \mathbf{S}_{i-1}, \mathbf{S}_{i}) \times\\
& \underbrace{\left( \sum\limits_{ \mathbf{U}_{i+1:N}} \prod_{w=i+1}^{N}\mathrm{Pr}(\mathbf{U}_{w})
 \mathrm{p}(\mathbf{Y}_{w}|\mathbf{X}_{w}) \mathrm{T}_{w}(\mathbf{U}_{w}, \mathbf{X}_{w}, \mathbf{S}_{w-1}, \mathbf{S}_{w})\right)}_{\triangleq\beta (\mathbf{S}_{i})},
\end{split}
\end{equation}
where the first part  behind "$\propto$"  denotes
the  forward-probabilities  $\alpha(\mathbf{S}_{i})$  and they are
 given by
\begin{equation}\label{alpha}
\begin{split}
\alpha(\mathbf{S}_{i})\propto
\sum_{\sim\mathbf{S}_{i}} \alpha(\mathbf{S}_{i-1}) \mathrm{Pr}(\mathbf{U}_{i})\mathrm{p}(\mathbf{Y}_{i}|\mathbf{X}_{i}) \mathrm{T}_{i}(\mathbf{U}_{i}, \mathbf{X}_{i}, \mathbf{S}_{i-1},\mathbf{S}_{i}).
\end{split}
\end{equation}
Similarly,  in (\ref{zuipai})   the third part behind "$\propto$"  represents the
backward-probabilities  $\beta(\mathbf{S}_{i})$,
 which  are  obtained  as
\begin{equation}\label{beta}
\begin{split}
&\beta(\mathbf{S}_{i})\propto \sum_{\sim\mathbf{S}_{i+1}}\\
&\beta(\mathbf{S}_{i+1}) \mathrm{Pr}(\mathbf{U}_{i+1})\mathrm{p}(\mathbf{Y}_{i+1}|\mathbf{X}_{i+1}) \mathrm{T}_{i+1}(\mathbf{U}_{i+1}, \mathbf{X}_{i+1}, \mathbf{S}_{i},\mathbf{S}_{i+1}).
\end{split}
\end{equation}
In summary,  the
 \emph{ a posteriori} information  in our   BP-MSDD  can be calculated by  using
\begin{equation}
\begin{split}\label{zui}
&\mathrm{Pr}(\mathbf{U}_{i}|\mathbf{Y})\propto \sum_{\sim \mathbf{U}_{i}}  \\
&\alpha(\mathbf{S}_{i-1})\beta(\mathbf{S}_{i})\mathrm{p}(\mathbf{Y}_{i}|\mathbf{X}_{i}) \mathrm{Pr}(\mathbf{U}_{i})\mathrm{T}_{i}(\mathbf{U}_{i}, \mathbf{X}_{i}, \mathbf{S}_{i-1},\mathbf{S}_{i}).
\end{split}
\end{equation}
Finally, the channel  decoder  calculates the decision statistics
$\imath(d_{i}(k))\triangleq \mathrm{p}(\mathbf{U}_{i}|\mathbf{Y})_{d_{i}(k)=1}  - \mathrm{p}(\mathbf{U}_{i}|\mathbf{Y})_{d_{i}(k)=0}$.
If  $\imath(d_{i}(k))\geq 0$,  we have    $\hat{d}_{i}(k)=1$.
Otherwise, we have  $\hat{d}_{i}(k)=0$.
\begin{figure}
 \begin{centering}
\includegraphics[width= 8 cm]{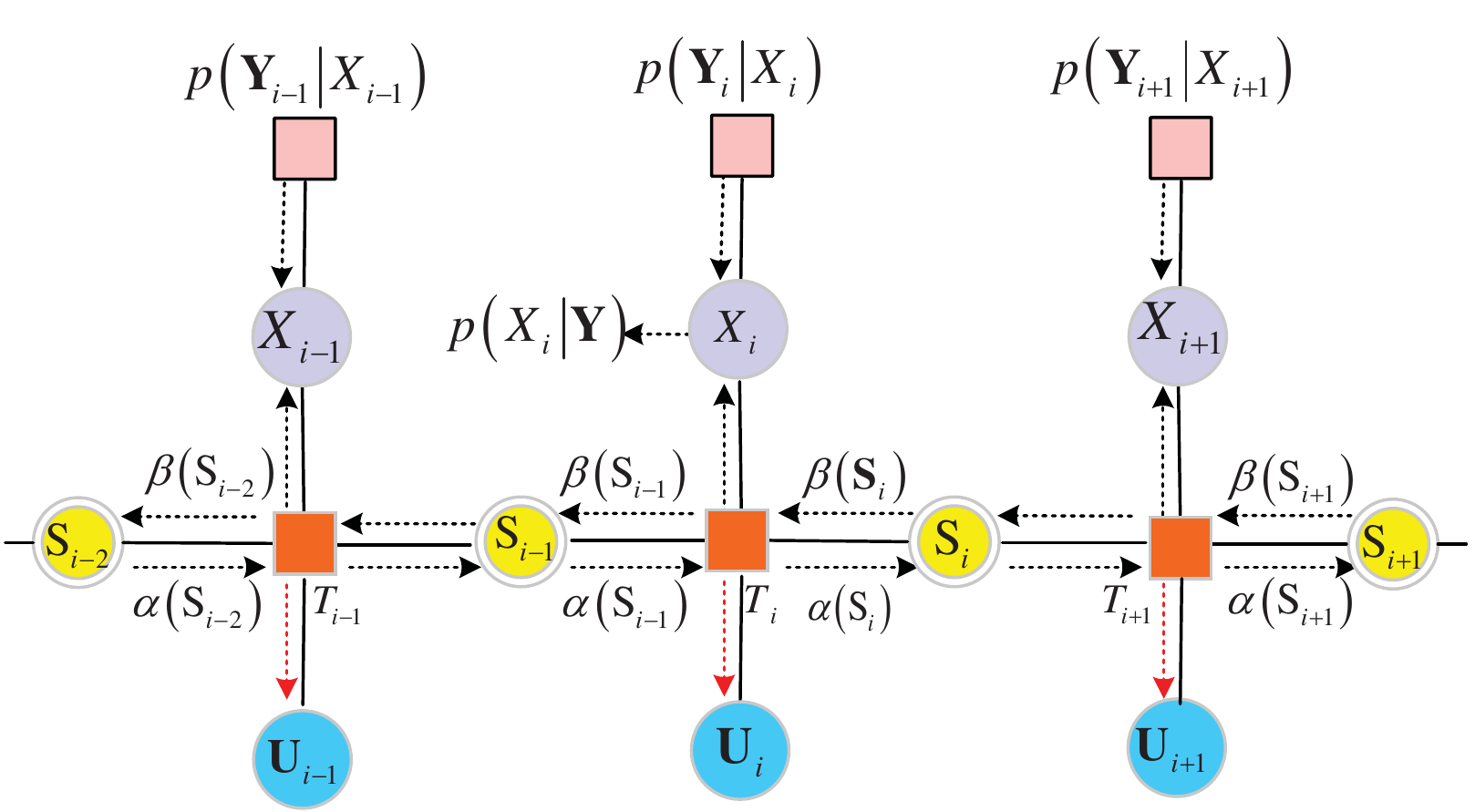}
\caption{\label{graph} The factor  graph representation of the proposed  BP-MSDD, where the larger  circles are the variable nodes corresponding to the  inputs and the  outputs,  the smaller  circles represent the  variable nodes corresponding to the  states, while the  squares are the  factor  nodes  characterizing the  check functions. Additionally, the  black arrows denote the directions of the  BP message  passing,
and the  red arrows denote the information sent from  the  detector to  the   decoder.}
\par\end{centering}
\end{figure}
In order to  thoroughly understand the proposed    BP-MSDD,
we employ a factor graph to model and visualize the  message passing mechanism of the BP-MSDD in the DSTBC aided UWB-IR  systems.
The factor graph  representation of the BP-MSDD  is shown in  Fig. \ref{graph},
where the larger  circles are the \emph{variable nodes} corresponding to the   inputs or the outputs,
the smaller   circles represent the \emph{variable nodes} corresponding to the   states,
while the  squares are the \emph{factor  nodes} characterizing the   \emph{check functions}.
More specifically, in the given factor graph each variable is represented by one variable node, each local function is represented by one factor node.
Additionally,
if and only if
a  variable is an argument of a  \emph{local function}, there exists an \textit{edge connecting the variable node and the factor node}.
The resultant BP-MSDD  has a forward-and-backward message passing mechanism.
To elaborate a little further,  the factor graph obtained  is a standard bipartite graph representing  the  relationship between  variables and local functions,
where each  local function  is essentially   the trellis check  function  $\mathrm{T}_{i}(\mathbf{U}_{i}, \mathbf{X}_{i}, \mathbf{S}_{i-1},\mathbf{S}_{i})$.
Note  that  in  BP-MSDD  we have  to calculate two  types of  recursion, i.e.,
the   forward-probabilities
 $\alpha(\mathbf{S}_{i})$, which  are  functions  of  $\alpha(\mathbf{S}_{i-1})$ and the likelihood message  $\mathrm{p}(\mathbf{Y}_{i}|\mathbf{X}_{i})$;
and  the   backward-probabilities
   $\beta(\mathbf{S}_{i})$,
    which are  functions  of $\beta(\mathbf{S}_{i+1})$ and $\mathrm{p}(\mathbf{Y}_{i+1}|\mathbf{X}_{i+1})$.
Finally,  the
 \emph{ a posteriori} information  are calculated  relying on   $ \alpha (\mathbf{S}_{i-1})$, $\beta (\mathbf{S}_{i})$  and  $\mathrm{p}(\mathbf{Y}_{i}|\mathbf{X}_{i})$ using (\ref{cond}).
In general,  the key operation is to calculate the
 \emph{ a posteriori} information of the  bits using  (\ref{zui}).
For the sake of  clarity, the detailed procedure for calculating the
 \emph{ a posteriori} information with the aid of the proposed BP-MSDD  is   summarized in Table I.

\begin{table}
\begin{center}
\caption{\label{tab1} Algorithm 1: Calculating (\ref{zui})  based  on the proposed BP-MSDD}
\begin{tabular}{ll}
\hline
Step 1) &  Initialize the forward-and-backward passed messages   as   \\
& $\alpha (\mathbf{S}_{0})=1$  and $\beta (\mathbf{S}_{0})=1$, respectively.\\
Step 2) &  Calculate  the branch  transition probabilities $\mathrm{p}(\mathbf{Y}_{i}|\mathbf{X}_{i})$  \\
       &  using (\ref{cond});  calculate
$\alpha (\mathbf{S}_{i})$ and  $\beta (\mathbf{S}_{i})$ with  (\ref{alpha}) and (\ref{beta}).\\
Step 3) &   Obtain the  \emph{ a posteriori} information  of the  bits using  (\ref{zui}),  \\
       &  which    is the key operation in the   proposed BP-MSDD   and \\
        &   relies  on   $\alpha (\mathbf{S}_{i})$,   $\beta (\mathbf{S}_{i})$ and $\mathrm{p}(\mathbf{Y}_{i}|\mathbf{X}_{i})$ calculated in Step 2).                  \\
\hline
\end{tabular}\\
\end{center}
\end{table}

\noindent\textbf{Remark 2:}
In  the  proposed  BP-MSDD,
"B" denotes the belief statistics  concerning   the information  symbols, which   are   inferred  using   the   \emph{ a posteriori} information
 obtained from  (\ref{zui}).
"P"  represents the  propagation process of the messages/belief statistics,  including  the
forward-probabilities
   $\alpha(\mathbf{S}_{i})$,
the backward-probabilities
$\beta(\mathbf{S}_{i+1})$,
the  likelihood  messages   $\mathrm{p}(\mathbf{Y}_{i}|\mathbf{X}_{i})$,
and the   \emph{ a posteriori} information    $\mathrm{Pr}(\mathbf{U}_{i}|\mathbf{Y})$.

\noindent\textbf{Remark 3:}
Note that  the  proposed BP-MSDD can be  employed in conjunction   with SISO channel decoding to construct    an  IDD receiver.
 When the proposed BP-MSDD  is applied  to  the context of the IDD receiver,
the  \emph{a priori}  probability becomes dynamic and   must be explicitly considered
during   the IDD   process, where soft-decision   information is exchanged between the BP-MSDD detector and the  decoder.

\section{Reduced-Complexity Alternatives to   BP-MSDD }

\subsection {  Transformation  of  Belief Propagation Based MSDD }

The proposed  BP-MSDD still imposes   high computational complexity, which increases   exponentially with the  observation window  size $M$.
To address this issue,  we first propose the  VA-HMSDD,
where the branch metric is regarded as  the approximate  decision  metric of the MSDD, resulting in an   observation interval shorter than $M$.
Bearing this in mind and inspired by the trellis-based detection, we further develop an iterative  VA-SMSDD  that serves as a reduced-complexity alternative to the high-complexity  iterative BP-MSDD in the context of  DSTBC aided UWB-IR systems.
With the aim of reducing the  computational  complexity of BP-MSDD,
we   resort to developing   a  simplified  expression  for  BP-MSDD  by invoking  some mathematical manipulations.
Specifically,  as far as the given information symbol is considered,
the log-likelihood ratio (LLR) of the bit $d_i(k) $   can be   expressed   as
\begin{equation}
\label{llr}
\ell\left(d_{i}(k)\right)=\mathrm{ln}\frac{\mathrm{Pr}(d_{i}(k)=1|\mathbf{Y})}{\mathrm{Pr}(d_{i}(k)=0|\mathbf{Y})},
\end{equation}
where both  the numerator and the  denominator are    APPs.
From the state transition  perspective,
let $\mathbf{S}_{i-1}=s'$,  $\mathbf{S}_{i}=s $. Then,    the APP  of $d_i(k)$ is  reformulated as
\begin{equation}
\mathrm{Pr}(d_{i}(k)|\mathbf{Y})\propto  \sum\limits_{ (s' \rightarrow s ): d_{i}(k)=z } \mathrm{Pr}(s',s,\mathbf{Y}),
\end{equation}
where
$(s' \rightarrow s ): d_{i}(k)=z$
denotes  that   the output  $ d_{i}(k)=z$  is obtained  in the transition  $(s' \rightarrow s )$.
Therefore,   (\ref{llr})  is    rewritten as
\begin{equation}
\begin{split}\label{llr1}
\ell\left(d_{i}(k)\right)\propto \quad \mathrm{ln}\frac{ \sum\limits_{ (s' \rightarrow s ): d_{i}(k)=1 } \mathrm{Pr}(s',s,\mathbf{Y})}{\sum\limits_{ (s' \rightarrow s ): d_{i}(k)=0 } \mathrm{Pr}(s',s,\mathbf{Y})},
\end{split}
\end{equation}
where  the  state transition  probability  from   $s' \rightarrow s $   is given by
\begin{equation}
\mathrm{Pr}(s',s,\mathbf{Y})
=\mathrm{Pr}(d_{i}(k))\alpha(s')\beta(s)\\
\times \mathrm{p}(\mathbf{Y}_{i}|\mathbf{X}_{i})\mathrm{ T}_{i}(\mathbf{U}_{i}, \mathbf{X}_{i}),
\end{equation}
and  $\mathrm{Pr}(d_{i}(k))$  denotes the  \emph{a priori}  probability of $d_{i}(k)$.
Similar to the scenario of APP \cite{con},  it can be seen that
the state transition  probability also   consists of both  \emph{a priori}  and extrinsic components.
Therefore,   (\ref{llr1}) can be  reformulated as
\begin{equation}\label{llrsum}
\begin{split}
&\ell(d_{i}(k))=\underbrace{\mathrm{ln}\frac{\mathrm{Pr}(d_{i}(k)=1)}{\mathrm{Pr}(d_{i}(k)=0)}}_{\triangleq \mathrm{\emph{a priori} \quad LLR} }  + \\
& \underbrace{\mathrm{ln}\frac{ \sum\limits_{ (s' \rightarrow s ): d_{i}(k)=1 } \alpha(s')\beta(s)\mathrm{p}(\mathbf{Y}_{i}|\mathbf{X}_{i}) \mathrm{T}_{i}(\mathbf{U}_{i}, \mathbf{X}_{i}, s', s)
}{\sum\limits_{ (s' \rightarrow s ): d_{i}(k)=0 } \alpha(s')\beta(s)\mathrm{p}(\mathbf{Y}_{i}|\mathbf{X}_{i}) \mathrm{T}_{i}(\mathbf{U}_{i}, \mathbf{X}_{i}, s', s)
}}_{ \triangleq \mathrm{extrinsic  \quad  LLR} =\ell^{\mathrm{e}}(d_{i}(k)) },
\end{split}
\end{equation}
where on the right-hand side  the first term  is   the  \emph{a priori}  LLR,
and    the second  term   represents the  extrinsic LLR  $\ell^{\mathrm{e}}(d_{i}(k))$.
It is noted that     $\ell^{\mathrm{e}}(d_{i}(k))$  includes  the   likelihood
$\mathrm{p}(\mathbf{Y}_{i}|\mathbf{X}_{i})$,
which   can be  approximately  obtained    from the     MSDD scheme  using   hard-decision.
In the following, the derivation of MSDD is revisited, based on which
the intrinsic connection  between the likelihood function and  the  MSDD  scheme is revealed.
More specifically, in the  $i$th
 observation window,  the   information-bearing symbols are  given by the set
$\mathbb{U}=[\mathbf{U}_{i-M+1},  \cdots, \mathbf{U}_{i}]$,
and the set of   differentially  encoded transmitted symbols   is  denoted as  $\mathbb{D}=[\mathbf{D}_{i-M}, \cdots, \mathbf{D}_{i}]$.
Assume that  $\mathbb{\tilde{U}}$ and $\mathbb{\tilde{D}(\tilde{U})}$ represent   the set of the estimated information symbols
and the set of the  estimated symbols   transmitted.
Then, at  the receiver,  the  GLRT-MSDD     boils down to
\begin{equation}\label{msddmetric2}
\mathbb{\hat{U}}=\mathrm{arg} \max \limits_{\mathbb{U}}\left[\Lambda\left(\{y_{q}(t)\}_{q=1}^{Q}|\mathbb{U}\right) \right]
\end{equation}
with   $\Lambda\left(\{y_{q}(t)\}_{q=1}^{Q}|\mathbb{U}\right)$  being the  likelihood function.
In addition,  we have
\begin{equation}\label{zhengbiyu}
\mathrm{p}(\mathbf{Y}_{i}|\mathbf{X}_{i}) \propto \mathrm{exp}\left (\Lambda\left(\{y_{q}(t)\}_{q=1}^{Q}|\mathbb{U}\right)\right).
\end{equation}
Since  $h_{p,q}(t)$    is unknown,  a reasonable alternative to  detect the  information symbols   is the  GLRT  algorithm \cite{tianzhi}.  Instead of directly estimating the information symbols in U,  the GLRT-MSDD maximizes the following log-likelihood function:
\begin{equation}\label{loglike}
\begin{split}
&\Lambda\left(\{y_{q}(t)\}_{q=1}^{Q}|\mathbb{U}\right)=\Lambda\left(y_{q}(t)|\mathbb{\tilde{D}(\tilde{U})},\tilde{g}_{p,q}(t)\right)  \\
& =2\int_0 ^ {MT_{\mathrm{s}}} y_{q}(t)\tilde{x}_{q}(t)dt
-\int_0 ^ {MT_{\mathrm{s}}} (\tilde{x}_{q}(t))^{2} dt,
\end{split}
\end{equation}
where   $\tilde{x}_{q}(t)$  is  the  \emph{candidate received signal} constructed by the  candidate symbol ${\tilde{d}}_{p,j}$ and  the  channel  template $\tilde{g}_{q,p}(t)$   according to
$\tilde{x}_{q}(t)=\sqrt{\frac{E_{\mathrm{b}}}{2}}\sum\limits_{p=1}^{2}\sum\limits_{j=1}^{2M}\tilde{d}_{p,j}\tilde{g}_{p,q}\left(t-(j-1)T_{\mathrm{f}}\right)+n_{q}(t)$,
and
$\tilde{g}_{p,q}(t)=\frac{1}{M\sqrt{2E_{\mathrm{b}}} }\sum\limits_{j=1}^{2M}\tilde{d}_{p,j}y_{q}\left(t+(j-1)T_{\mathrm{f}}\right)$ \cite{wtt}.
Substituting  $\tilde{x}_{q}(t)$        and     $\tilde{g}_{q,p}(t)$   into (\ref{loglike}),   we obtain
\begin{equation}\label{lam}
\begin {split}
&\Lambda\left(\{y_{q}(t)\}_{q=1}^{Q}|\mathbb{U}\right)
=\sum\limits_{p=1}^{2}\Lambda\left(y_{q}(t)|\{\tilde{d}_{p,j}\},\tilde{g}_{p,q}(t)\right)  \\
&=\sum\limits_{p=1}^{2}  \int_{0}^{T_{\mathrm{i}}}\left(\sum\limits_{j=1}^{2M}\tilde{d}_{p,j}y_{q}(t+(j-1)T_{\mathrm{f}})\right)^{2}dt.
\end{split}
\end{equation}
For convenience of expression,     (\ref{lam})  is    reformulated  in the matrix form   as
\begin{equation}\label{msddmetric1}
\begin{split}
&\Lambda\left(\{y_{q}(t)\}_{q=1}^{Q}|\mathbb{U}\right)= \sum\limits_{i'=1}^{M}\sum\limits_{l=0}^{i'-1}\mathrm{Tr}\left(\prod_{v=-i'+1}^{-l}\mathbf{{\tilde{U}}}_{i+v}\right)  \\
& \times  \left(\sum\limits_{q=1}^{Q}\int_{0}^{T_{\mathrm{i}}}\left(\mathbf{y}_{i-l}^{q}(t)\right)^{\mathrm{T}}\mathbf{y}_{i-i'}^{q}(t)dt\right)\\
&  =\sum\limits_{i'=1}^{M}\sum\limits_{l=0}^{i'-1}\mathrm{Tr}\left(\left(\prod_{v=-i'+1}^{-l}\mathbf{{\tilde{U}}}_{i+v}\right)\left(\sum\limits_{q=1}^{Q} \mathbf{Y}^{q}_{i-i',l-i'}\right)\right).
\end{split}
\end{equation}
Based on (\ref{zhengbiyu}), (\ref{loglike})    and (\ref{msddmetric1}),  the likelihood $\mathrm{p}(\mathbf{Y}_{i}|\mathbf{X}_{i})$ is given by
\begin{equation}\label{zhengbi}
\begin{split}
&\mathrm{p}(\mathbf{Y}_{i}|\mathbf{X}_{i}) \propto \mathrm{exp}  \\
&\left\{\sum\limits_{i'=1}^{M}\sum\limits_{l=0}^{i'-1}\mathrm{Tr}\left(\left(\prod_{v=-i'+1}^{-l}\mathbf{{\tilde{U}}}_{i+v}\right)\left(\sum\limits_{q=1}^{Q} \mathbf{Y}^{q}_{i-i',l-i'}\right)\right)\right\}.
\end{split}
\end{equation}
Substituting  $\mathrm{p}(\mathbf{Y}_{i}|\mathbf{X}_{i})$ into  (\ref{llrsum}),
the  LLR  $\ell(d_{i}(k))$    can be calculated.
\noindent\textbf{Remark 4:}
Note that the LLR-based  BP-MSDD   can be  implemented using (\ref{llrsum}),
and the GLRT-MSDD  is   expressed   as  (\ref{msddmetric1}).
However, both of them face a critical challenge in certain practical applications, since
the computational complexity   of   calculating    (\ref{llrsum}) and (\ref{msddmetric1})
increases    exponentially   with $M$.
Therefore,  in the following, a pair of reduced-complexity MSDD schemes, namely the    VA-HMSDD  and the VA-SMSDD are proposed
to simplify the classic  GLRT-MSDD   and  the    BP-MSDD, respectively.

\subsection {Viterbi Algorithm Based  Hard-Decision MSDD}

\begin{figure}
 \begin{centering}
\includegraphics[width= 7.5 cm]{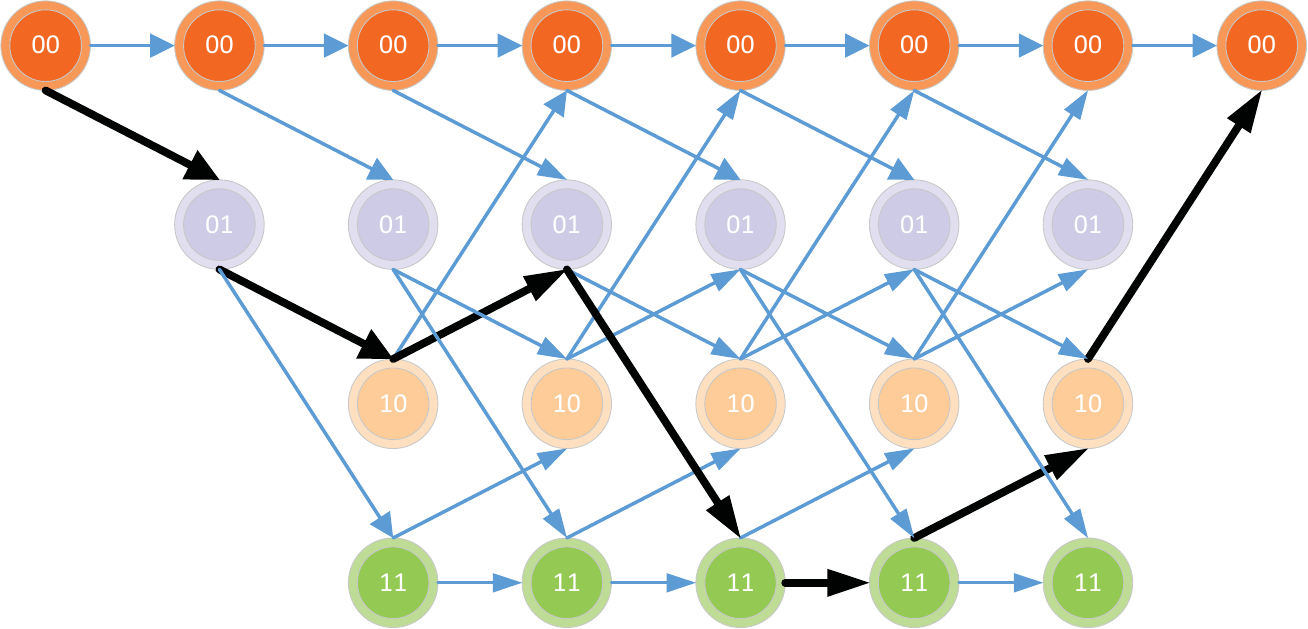}
\caption{\label{vatrellis} An illustration of the  VA trellis   corresponding to the given   input sequence \{1011100\},
where the
data in the
circles represent the states,  the  arrows   represent all the possible paths, while the black bold arrows  represent  the survivor path having covered all the states.  The survivor path  is obtained in accordance with the selection criteria of VA.}
\par\end{centering}
\end{figure}

Upon employing   the principle of reduced-state trellis detection  \cite{lowcom},
a survivor   path can be constructed   in   the MSDD trellis diagram based on VA.
In particular, the computational complexity  of  the    likelihood function    (\ref{msddmetric1})  can be  specified   with a fixed memory depth of   $L $ $(L\leq M) $, which  is  related to the range of   $L $   and is
a  key  design parameter  determining   the computational  complexity.
In order to make the VA applicable to  MSDD,
we attempt to construct a simplified  expression of   GLRT-MSDD, whose expression     $\Lambda\left(\{y_{q}(t)\}_{q=1}^{Q}|\mathbb{U}\right)$  is approximated as
\begin{equation}\label{va}
\begin{split}
&\Gamma\left(\{y_{q}(t)\}_{q=1}^{Q}|\mathbb{U}\right)=
\sum\limits_{i'=1}^{M}\sum\limits_{l=\mathrm{max}\{0,i'-L\}}^{i'-1}  \\
&\mathrm{Tr}\left(\left(\prod_{v=-i'+1}^{-l}\mathbf{{\tilde{U}}}_{i+v}\right)\left(\sum\limits_{q=1}^{Q} \mathbf{Y}^{q}_{i-i',l-i'}\right)\right).
\end{split}
\end{equation}
In  (\ref{va}), the number of   addends  indexed by $l$  entails a maximum of $L$   instead of    $i'$.
Thus,  the  approximate   expression    $\Gamma\left(\{y_{q}(t)\}_{q=1}^{Q}|\mathbb{U}\right)$  for MSDD
 is  constrained by   a fixed memory depth  $L$,
which   facilitates     the VA implementation.
Let us set
$ \mathbf{{\tilde{U}}}_{i}^{(L)} \triangleq [ \mathbf{\tilde{U}}_{i-L},  \cdots, \mathbf{\tilde{U}}_{i}]^{\mathrm{T}}$, then  (\ref{va}) is  rewritten as
\begin{equation}\label{val}
\Gamma\left(\{y_{q}(t)\}_{q=1}^{Q}|\mathbf{{\tilde{U}}}_{i}^{M}\right)
=\sum\limits_{i'=1}^{M}\lambda_{i'}(\mathbf{{\tilde{U}}}_{i-1}^{(L-1)},\mathbf{{\tilde{U}}}_{i}),
\end{equation}
where we have
\begin{equation}\label{lambda}
\begin{split}
&\lambda_{i'}(\mathbf{{\tilde{U}}}_{i-1}^{(L-1)},\mathbf{{\tilde{U}}}_{i})\triangleq  \\
&\sum\limits_{l=\mathrm{max}\{0,i'-L\}}^{i'-1}\mathrm{Tr}\left(\left(\prod_{v=-i'+1}^{-l}\mathbf{{\tilde{U}}}_{i+v}\right)\left(\sum\limits_{q=1}^{Q} \mathbf{Y}^{q}_{i-i',l-i'}\right)\right).
\end{split}
\end{equation}
Note that  with a trellis representation, (\ref{lambda}) stands for the branch metric  at the
$i'$th stage in a trellis diagram,
and it  depends on the trellis  $\mathbf{{\tilde{U}}}_{i}^{(L)}$  only.
With  $J_{i'}(\mathbf{{\tilde{U}}}_{i}^{M})$ denoting  the accumulated metric  at the
$i'$th stage,
it is plausible  that  the accumulated metric until the current stage
can be formulated as the sum of that at the previous stage plus the current branch metric.
More explicitly, we have
\begin{equation}\label{zongdaijia}
J_{i'}(\mathbf{{\tilde{U}}}_{i}^{M})=J_{i'-1}(\mathbf{{\tilde{U}}}_{i-1}^{M-1})+ \lambda_{i'}(\mathbf{{\tilde{U}}}_{i-1}^{(L-1)},\mathbf{{\tilde{U}}}_{i}),
\end{equation}
which is an equivalent formulation to (\ref{va}).
As a beneficial result,  the     VA-HMSDD   has been   constructed.
According to  (\ref{lambda}) and (\ref{zongdaijia}), the sequence of states  corresponding to the survivor path
yields  the maximum value of the  accumulated metric   and  uniquely defines the winning  sequence of trellis state transitions   from    the first stage
to   the last one.   As a result,   the transmitted  symbol sequence  can be    decided.
Let us consider  the given   input sequence \{1011100\}  as an example.
Specifically, in   Fig. \ref{vatrellis}    the hard-decision VA trellis is plotted,
where  the data in
 the circles represent the states,
the arrows  represent all the possible paths,
while the black bold arrows    represent  the survivor path states.
The  survivor path  is obtained   in accordance with the selection criteria of VA.
For the sake of  clarity,  the    detailed procedure of the VA-HMSDD    is   specified  in   Table II.
Note    that  the computational complexity  of VA-HMSDD  is fixed   due to    a  given  memory depth  of $L$.
This  fulfills  our   goal of reducing the  complexity of the conventional GLRT-MSDD,
because we have   $L\leq M$.
Moreover, the proposed VA-HMSDD  achieves    the same detection performance as the  GLRT-MSDD   when $L= M$.
In particular, when $M=1$, the complexity of the  GLRT-MSDD cannot be simplified with the VA-HMSDD.
In order to further enhance   the detection performance of VA-HMSDD,
in the next subsection,  we develop  a   VA-SMSDD scheme, which is obtained by extending the trellis-based detection philosophy to the  DSTBC  aided UWB-IR systems.

\begin{table}
\begin{center}
\caption{\label{tab1} Algorithm 2: The implementation  of  VA-HMSDD }
\begin{tabular}{ll}
\hline
Step 1) &   Start from the initial state, and  set  $J_{0}=0$;\\
Step 2) &   Set
$\lambda_{1}(\mathbf{{\tilde{U}}}_{1})=\mathbf{Y}_{0,1}\mathbf{{\tilde{U}}}_{1}$, and update  the  accumulated
           function   \\
           & using  $ J_{1}(\mathbf{{\tilde{U}}}_{1})=J_{0} + \lambda_{1}(\mathbf{{\tilde{U}}}_{1})=\mathbf{Y}_{0,1}\mathbf{{\tilde{U}}}_{1}$; \\
Step 3) &   Set
$\lambda_{2}(\mathbf{{\tilde{U}}}_{1},\mathbf{{\tilde{U}}}_{2})= \mathbf{Y}_{1,2}\mathbf{{\tilde{U}}}_{2}+
\mathbf{Y}_{0,2}\mathbf{{\tilde{U}}}_{1}\mathbf{{\tilde{U}}}_{2}$   and  update   the  \\
& accumulated   function for VA-HMSDD  using  \\
&   $J_{2}(\mathbf{{\tilde{U}}}_{2})= J_{1}(\mathbf{{\tilde{U}}}_{1})+ \lambda_{2}(\mathbf{{\tilde{U}}}_{1},\mathbf{{\tilde{U}}}_{2}) $ \\
 &    $=\mathbf{Y}_{0,1}\mathbf{{\tilde{U}}}_{1}+  \mathbf{Y}_{1,2}\mathbf{{\tilde{U}}}_{2}+$
$\mathbf{Y}_{0,2}\mathbf{{\tilde{U}}}_{1}\mathbf{{\tilde{U}}}_{2}$;\\
Step  $\cdots$ &  \\
Step $M$) &  Set  and update the accumulated function  using \\
&  $\Gamma\left(\{y_{q}(t)\}_{q=1}^{Q}|\mathbb{U}\right)=J_{M}(\mathbf{{\tilde{U}}}_{M})$  \\
& $=J_{M-1}(\mathbf{{\tilde{U}}}_{i-1}^{{M-1}})+\lambda_{M}(\mathbf{{\tilde{U}}}_{i-1}^{(L-1)},
\mathbf{{\tilde{U}}}_{M})$. \\
\hline
\end{tabular}\\
\end{center}
\end{table}

\subsection {Viterbi Algorithm Based  Soft-Decision MSDD}

The accumulated    metric  of  (\ref{val})  is related   to   the current input information symbol and  the input information symbols of the previous stages.
Consequently,  the accumulated metric  can be implemented  on the trellis diagram.
Upon invoking  some mathematical manipulations  concerning   (\ref{val}),
we can explicitly    formulate    the  forward-and-backward branch metrics  as
\begin{equation}
\lambda( \alpha_{i-1} \rightarrow \alpha_{i} )=\sum\limits_{i'=1}^{M}\lambda_{i'}(\mathbf{{\tilde{U}}}_{i-L}^{(L-1)},\mathbf{{\tilde{U}}}_{i})
\end{equation}
and
\begin{equation}
\lambda( \beta_{i+1} \rightarrow \beta_{i} )=\sum\limits_{i'=1}^{M}\lambda_{i'}(\mathbf{{\tilde{U}}}_{i+L}^{(L-1)},\mathbf{{\tilde{U}}}_{i+1}),
\end{equation}
respectively.
Correspondingly,  relying on the branch metrics,
the forward-and-backward accumulated   path metrics are  given by
\begin{equation}\label{uf}
u_{\mathrm{forward}}(\alpha_{i})= \max \limits_{\{\alpha_{i-1}\}}\{u_{\mathrm{forward}}(\alpha_{i-1})+\lambda( \alpha_{i-1} \rightarrow \alpha_{i} )\}
\end{equation}
and
\begin{equation}\label{ub}
u_{\mathrm{backward}}(\beta_{i})= \max \limits_{\{\beta_{i+1}\}}\{u_{\mathrm{backward}}(\beta_{i+1})+\lambda( \beta_{i+1} \rightarrow \beta_{i} )\},
\end{equation}
respectively.
Based on     (\ref{uf}) and (\ref{ub}),
the   VA-SMSDD    is expressed  as
\begin{equation}\label{soft}
\begin{split}
\iota(d_{i}(k))&=\sum\limits_{(s' \rightarrow s ): d_{i}(k)=1}\{u_{\mathrm{forward}}(\alpha_{i})+u_{\mathrm{backward}}(\beta_{i})\}\\
&-\sum\limits_{(s' \rightarrow s ): d_{i}(k)=0}\{u_{\mathrm{fordward}}(\alpha_{i})+u_{\mathrm{backward}}(\beta_{i})\}.
\end{split}
\end{equation}
According to  the expression $\iota(d_{i}(k))$, the channel  decoder makes the decision as follows.
If  $\iota(d_{i}(k))\geq 0$,  then $\hat{d}_{i}(k)=1$;
otherwise,   $\hat{d}_{i}(k)=0$.
The proposed VA-SMSDD can be integrated into   an IDD receiver    when   the  \emph{a priori}  probabilities are
taken into account in   the detection process.
As a result, in  the
$r$th
 iteration between the detector and the channel decoder,
the accumulated
path metrics of (\ref{uf}) and (\ref{ub})  are   modified    as
\begin{equation}
\begin{split}
&u_{\mathrm{forward}}^{(r)}(\alpha_{i})=\max \limits_{\{\alpha_{i-1}\}}\{u_{\mathrm{forward}}^{(r)}(\alpha_{i-1}) \\
&+\lambda( \alpha_{i-1} \rightarrow \alpha_{i} )
+ \mathrm{ln}( \mathrm{Pr}^{(r-1)}(\alpha_{i-1}))\}
\end{split}
\end{equation}
and
\begin{equation}
\begin{split}
&u_{\mathrm{backward}}^{(r)}(\beta_{i})= \max \limits_{\{\beta_{i+1}\}}\{u_{\mathrm{backward}}^{(r)}(\beta_{i+1})\\
&+\lambda( \beta_{i+1} \rightarrow \beta_{i} )
+ \mathrm{ln} (\mathrm{Pr}^{(r-1)} (\beta_{i+1}))  \},
\end{split}
\end{equation}
respectively,  where $\mathrm{Pr}^{(r-1)} (\cdot)$ is   the   \emph{a priori}  probability  gleaned    from  the  channel  decoder in the previous iteration.

\noindent\textbf{Remark 5:}
By extending the  trellis-based detection philosophy,
the   iterative VA-SMSDD is   obtained, where  the
\emph{a priori}  probabilities are    incorporated into  the IDD process
for achieving a more competitive performance than
that   of VA-HMSDD.
Explicitly, the    VA-SMSDD  employs (\ref{soft}), which
facilitates    more reliable detection  than   the   VA-HMSDD  using   (\ref{zongdaijia}).

\begin{table}
\begin{center}
\caption{\label{tab1}  Computational   Complexity Comparisons  of Four MSDD Algorithms.}
\begin{tabular}{ll}
\hline
 Schemes     &   Computational  Complexity  \\
 \hline
\end{tabular}\\
\begin{tabular}{ll}
\hline
GLRT-MSDD    &   $O[(M+1)\cdot (2^{M})]$ \\
VA-HMSDD &  $O[(L+1) \cdot (2^{L})]$ \\
BP-MSDD  &  $O[M \cdot(M+1) \cdot 2^{M}]$\\
VA-SMSDD &  $O[M \cdot(L+1) \cdot 2^{L}]$ \\
 \hline
\end{tabular}\\
\end{center}
\end{table}

\section{Computational Complexity Analysis}

In this  section, the computational  complexity  of the conventional hard-decision GLRT-MSDD,
of the   VA-HMSDD,  of the  BP-MSDD and of the  VA-SMSDD are   compared.

Firstly, the complexity of the conventional  GLRT-MSDD with  hard-decision  is  analyzed.
The lattice structure in (\ref{msddmetric1})  results in an exhaustive evaluation of  all cross-correlation combinations.
Then,  the  hard-decision  GLRT-MSDD  can be implemented relying on    (\ref{msddmetric2}).
Therefore, in terms of the number of multiplications required, the   GLRT-MSDD
imposes a    computational  complexity of $O[(M+1)\cdot (2^{M})]$, i.e.,  it increases
exponentially with the observation   window size $M$.
By comparison,    for the  VA-HMSDD, as    seen    from  (\ref{lambda}),
only those cross-terms with a memory depth  no more than $L$ are taken into account.
Therefore,  the complexity of the VA-HMSDD  increases  exponentially with the memory depth $L$.
More specifically,   the  computational   complexity  of   VA-HMSDD   reduces to    $O[(L+1) \cdot (2^{L})]$ in terms of the total multiplications required.
Consequently, when   $L $  gets  larger,  the detection performance of the VA-HMSDD becomes closer to that of the  conventional hard-decision GLRT-MSDD.
Similar    conclusions are also valid   for
the   soft-decision  based  BP-MSDD  versus the VA-SMSDD.

Secondly,  the complexity of the proposed  BP-MSDD and  VA-SMSDD  is   compared.
Specifically, the    expression  of the   BP-MSDD given  in (\ref{zui})  imposes
a computational   complexity  of  $O[M \cdot(M+1) \cdot 2^{M}]$.
This  is  practically applicable   only when  $M$ is small.
In order to maintain an  affordable complexity  when $M$ increases,
VA-SMSDD is additionally conceived,
which  strikes   an appealing performance-versus-complexity tradeoff.
The  maximum affordable complexity   can  be imposed   through a proper selection of the  finite memory depth $L$.
As a result,   the computational   complexity  of   VA-SMSDD  is  on  the order of   $O[M \cdot(L+1) \cdot 2^{L}]$,
where  $2^{L}$  is  the  number of trellis states in each stage.
In other words,  its   complexity  increases exponentially only with $L$, rather than  with  $M$.
Therefore, both the VA-HMSDD and the VA-SMSDD enjoy a flexibility in striking specific performance-versus-complexity tradeoffs.
In summary,   BP-MSDD achieves the best   BER performance at the expense of    high  complexity.
By comparison,   VA-SMSDD  constitutes   a low-complexity detector   suffering    minor  performance loss.
For clarity, the  complexity comparisons  of   these four  MSDD  algorithms    are  summarized    in Table III.
It is observed that VA-HMSDD has a lower complexity than  the conventional hard-decision GLRT-MSDD,
while the complexity of  VA-SMSDD sits   between  that of  VA-HMSDD  and  that of  BP-MSDD.

\section{Simulation Results and Discussions }

\begin{figure}
 \begin{centering}
\includegraphics[width=7.5cm]{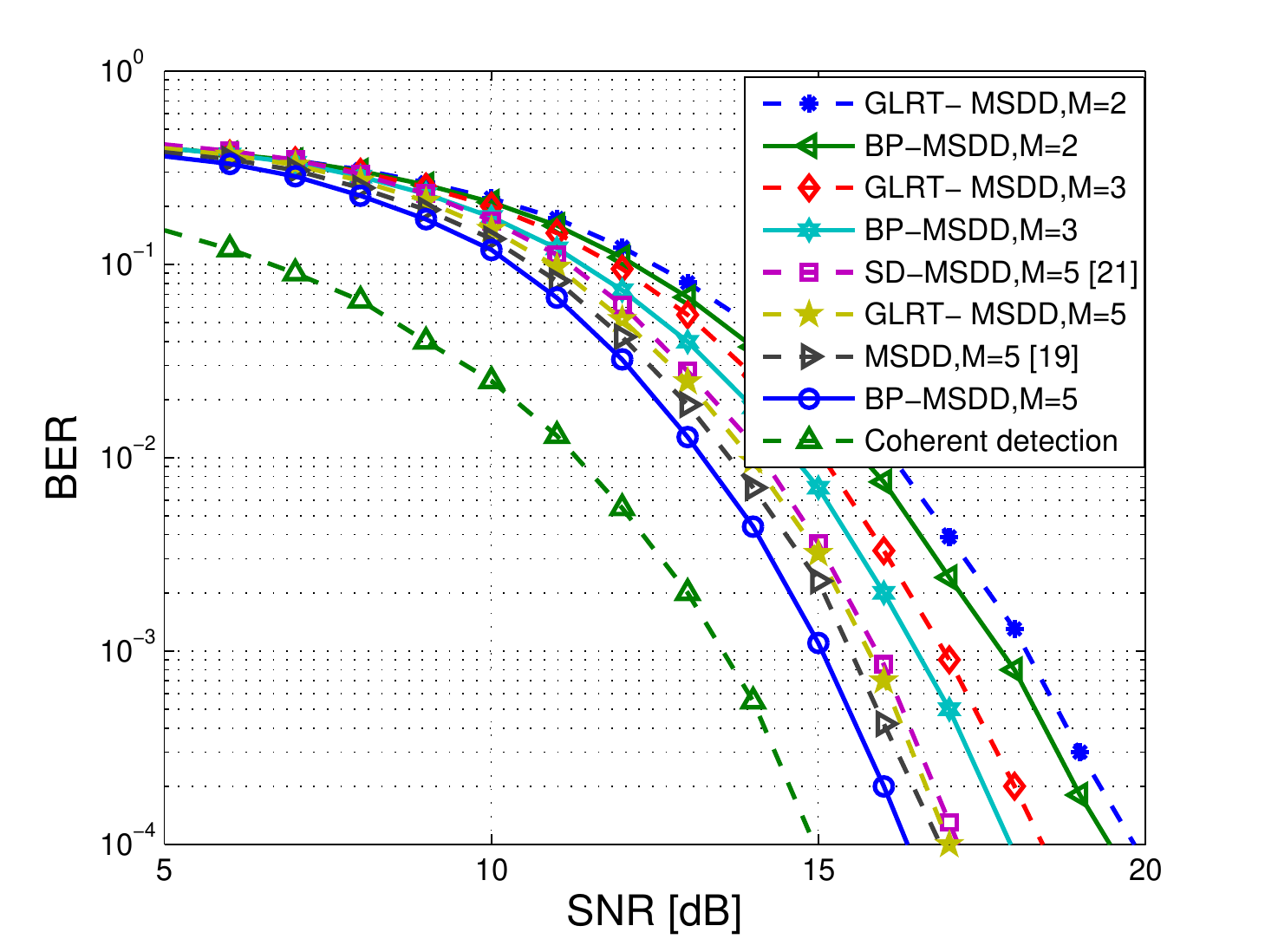}
\caption{\label{fig1a} BER performance comparisons of the coherent detector, the proposed  BP-MSDD, the conventional GLRT-MSDD  and the MSDDs proposed in  [19] and [21].
 $ Q = 1$, $M=2$, 3 and 5.}
\par\end{centering}
\end{figure}

\begin{figure}
\begin{centering}
\includegraphics[width=7.5cm]{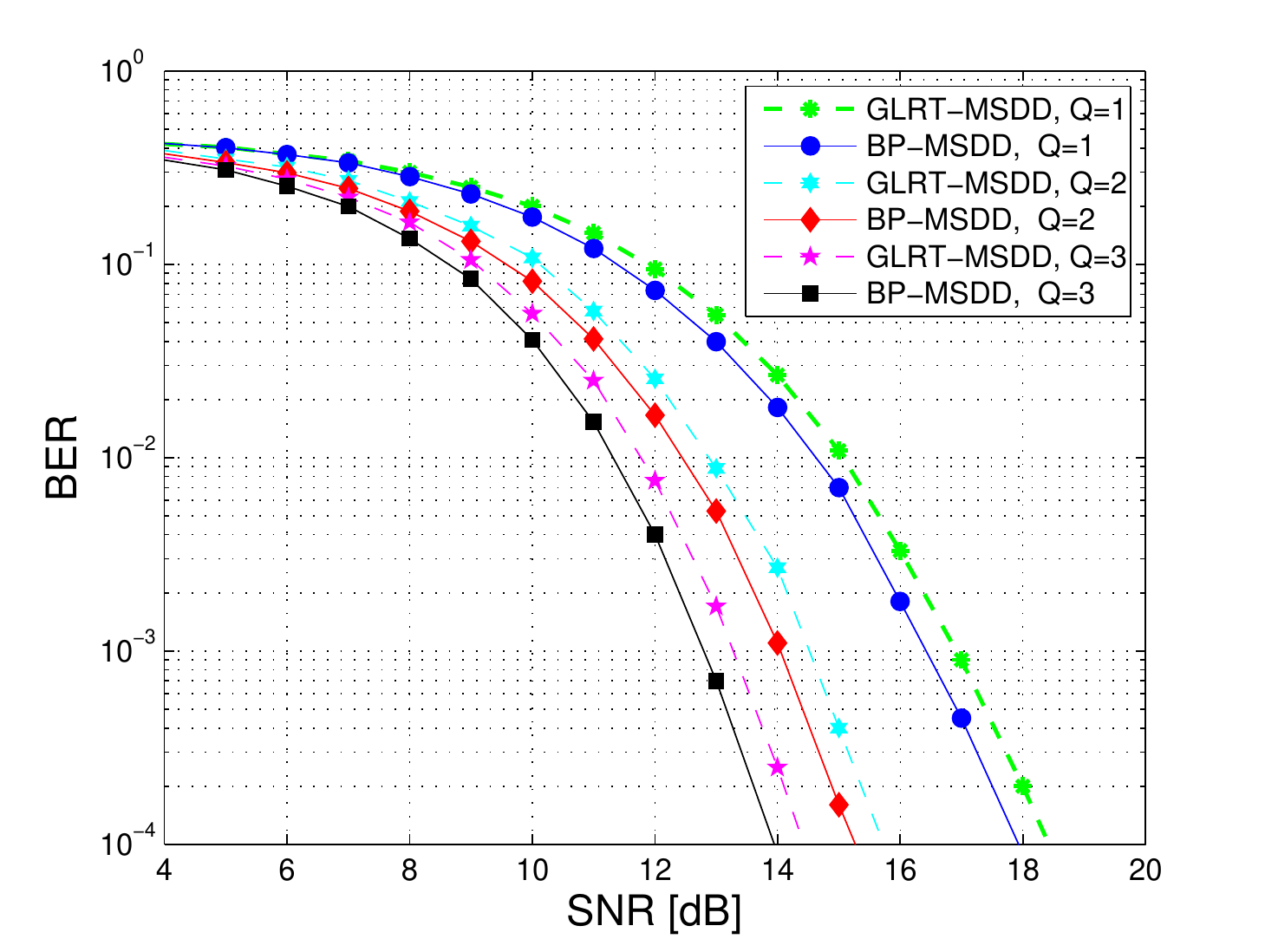}
\caption{\label{fig1b}BER performance comparisons  of the proposed   BP-MSDD and  the  conventional hard-decision GLRT-MSDD.  $M = 3$, $Q=1$, 2 and 3.}
\par
\end{centering}
\end{figure}

\begin{figure}
\begin{centering}
\includegraphics[width=7.5 cm]{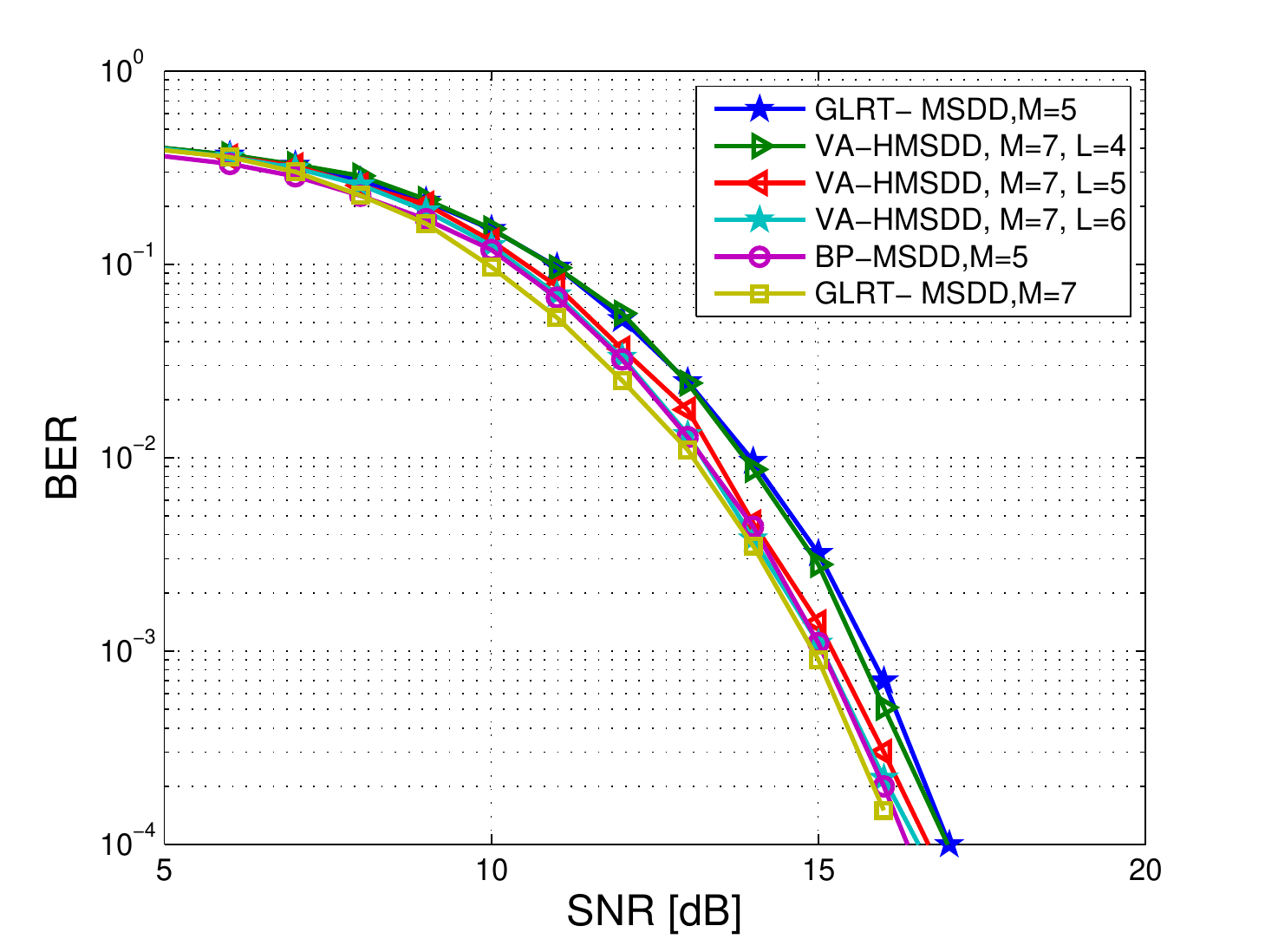}
\caption{\label{vahmsdd}BER performance comparisons  of the proposed   VA-HMSDD,   the conventional hard-decision   GLRT-MSDD, and the proposed BP-MSDD.
The  observation window size is $M = 5$, 7, while the value of $L$ varies.
}\par
\end{centering}
\end{figure}

\begin{figure}
\begin{centering}
\includegraphics[width=7.5cm]{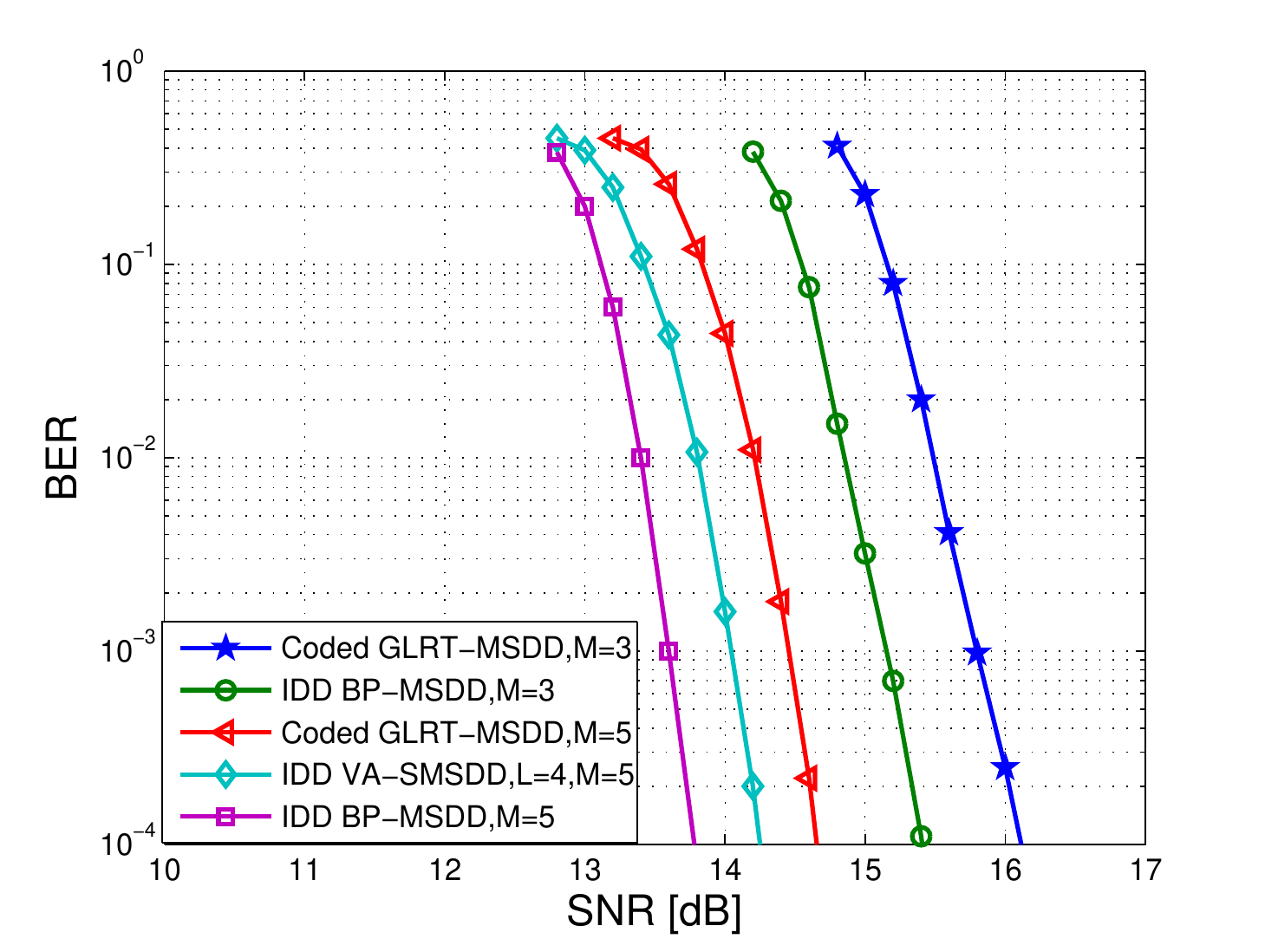}
\caption{\label{Fig10IDD} BER performance  comparisons of  the coded  GLRT-MSDD   with  $M = 3$, 5,  the IDD BP-MSDD  with    $M = 3$, 5  and the IDD VA-SMSDD     with   $L=4$,  $M = 5$  for $Q = 1$.}
\par\end{centering}
\end{figure}

\begin{figure}
\begin{centering}
\includegraphics[width=7.5cm]{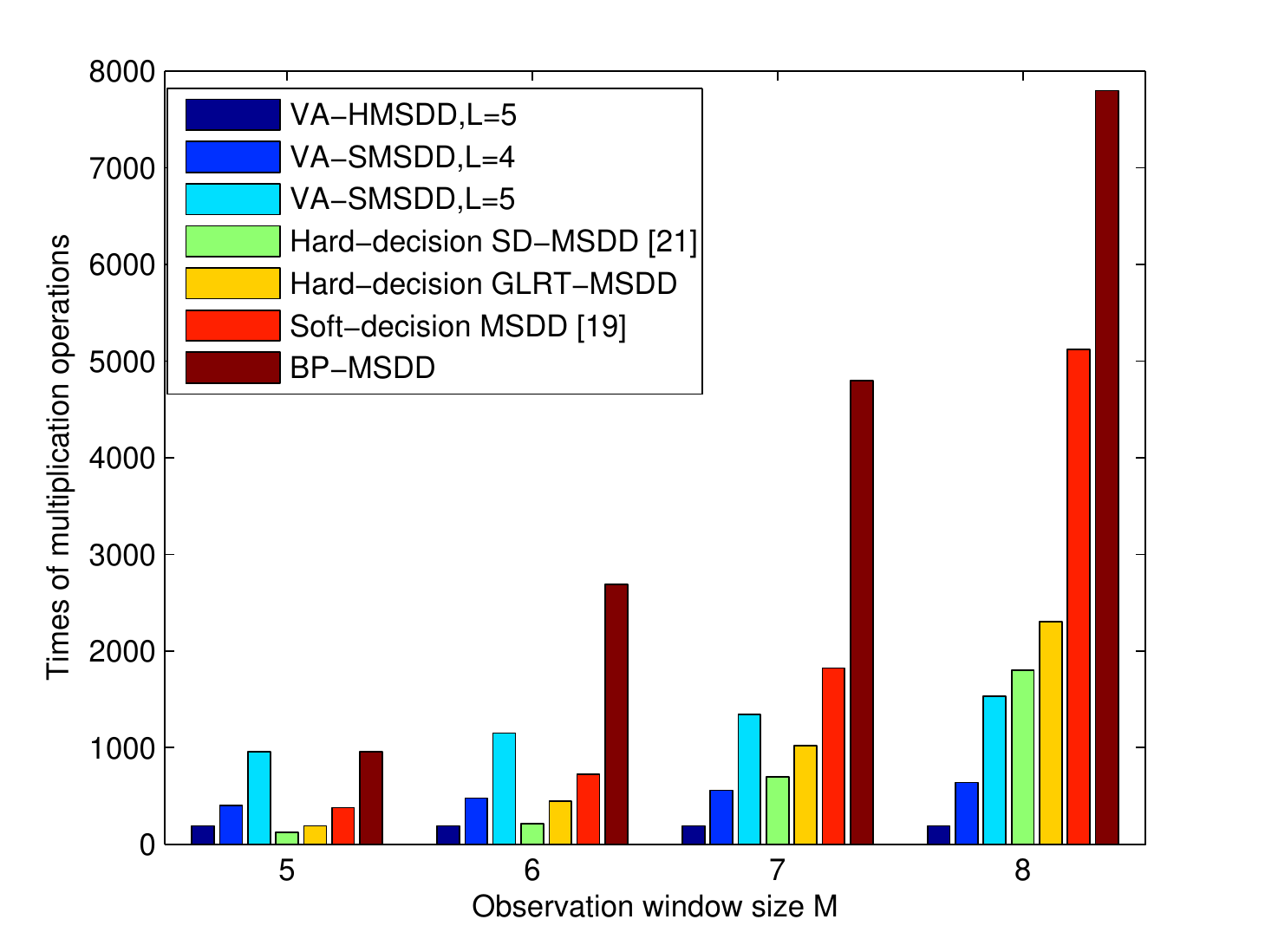}
\caption{\label{complexity} Complexity comparisons  of  the BP-MSDD, the soft-decision MSDD in  \cite{pisit},  the hard-decision GLRT-MSDD,  the hard-decision SD-MSDD in \cite{llr}, the VA-SMSDD and the VA-HMSDD  against different   observation window size $M$  and   different VA memory depth  $L$.
 }
\par
\end{centering}
\end{figure}

In this section, numerical  simulations are conducted to validate  the effectiveness of the proposed MSDD  algorithms.
In all the simulations,   a DSTBC aided  UWB-IR system  is considered.
The channel is generated based on the IEEE 802.15.3a CM2 model \cite{zq}.
This multipath channel remains   invariant during  each symbol burst, but randomly varies from burst to burst  \cite{win}.
The assumption of channel invariance during   each symbol burst is  reasonable,
since each   transmitted burst includes only   tens of symbols.
The monocycle waveform $\omega(t)$ employed  is the normalized second derivative of a Gaussian function, i.e.,
$ \omega(t)=[1-4\pi(t/ T_{\mathrm{m}})^{2}]\mathrm{exp}[-2\pi(t/T_{\mathrm{m}})^{2}]$,
where the pulse duration is
$ T_{\mathrm{m}}= 0.287 \mathrm{ns} $.
In order to eliminate the intersymbol interference, the frame duration is chosen as
$T_{\mathrm{f}}=80 \mathrm{ns}  $,
 so that we have
$T_{\mathrm{f}}$ $>$   $T_{\mathrm{n}}$,
 where
$T_{\mathrm{n}}=40  \mathrm{ns}  $,
   denotes the maximum excess delay of the channel.
Furthermore, the thermal noise is modeled as a  white Gaussian process, with the two-sided power spectral density being
$N_{0}/2$.
Firstly,    the  BER  performance of the proposed  BP-MSDD  is evaluated  in Fig.\ref{fig1a}  and Fig. \ref{fig1b},  where either
the number of   receive antennas $Q$ or the observation window size $M$ varies.
To obtain insights concerning  practical implementation,  VA-HMSDD is investigated   under different   values   of  the design parameter $L$ in Fig. \ref{vahmsdd}.
Furthermore,  the  BER performance  of  the  iterative BP-MSDD and of the iterative VA-SMSDD  is   compared in Fig. \ref{Fig10IDD}.
Finally,  the  computational    complexity comparisons    of the  BP-MSDD,  VA-SMSDD,  GLRT-MSDD and  VA-HMSDD   are  visualized in Fig. \ref{complexity}.

\emph{Test Scenario 1}:
This test scenario   is devoted to demonstrating the  effectiveness of the  proposed BP-MSDD  without considering the channel coding.
Specifically, in  Fig. \ref{fig1a} we  illustrate the BER performance comparisons of  the proposed  BP-MSDD, the conventional
GLRT-MSDD,
and  the MSDDs  proposed in   \cite{pisit} and \cite{llr}.
Moreover, a coherent  rake
 receiver \cite{Rake} with   12 fingers
is also evaluated as the performance benchmark.
The number of  receive antennas  is fixed to  $Q=1$, while  the   observation window size  is set to  $M=2, 3$  and  5, respectively.
It is clear  that   BP-MSDD  achieves   a better BER performance than the hard-decision
 GLRT-MSDD, e.g., about  0.4$\mathrm{dB}$ gain
is attained   at the BER of $10^{-4}$   when  $M=2$.
In addition, this   BER performance   advantage   becomes larger when  $M$ increases.
It can be seen that BP-MSDD  outperforms the   BCJR-based  MSDDs of   \cite{pisit}   when   $M=5$.
This is due to the more reliable information
obtained by  the BP-MSDD relying on
the proposed AcR  sampling and metric function than that obtained by the BCJR-based MSDD of \cite{pisit}.
Also, the BP-MSDD is superior to the  MSDD of \cite{llr}.
Although performance loss is observed in comparison with the coherent reception, the benefit of the proposed noncoherent BP-MSDD is still evident without channel estimation.
Furthermore,  the  BER    performance of  the proposed   BP-MSDD  is  investigated
by varying the number of  receive   antennas  $Q$.
Specifically, in   Fig. \ref{fig1b}   the BER performance of BP-MSDD  is plotted for the scenario where   $M=3$,    $Q=1$,  2  and 3,  respectively.
It is   observed that the curves in   Fig. \ref{fig1b}  have   similar trend  to  those of   Fig. \ref{fig1a},
and  in  Fig. \ref{fig1b}    the proposed BP-MSDD always enjoys a   performance improvement
 compared to the conventional hard-decision   GLRT-MSDD,
  regardless of the specific values of  $Q$.
This is attributed to   the advantage of the forward-and-backward message passing  mechanism  of the  BP-MSDD.

\emph{Test   Scenario   2}:
Compared  to  the   BP-MSDD and   the conventional hard-decision  GLRT-MSDD,
 the proposed VA-HMSDD   represents a suboptimal reduced-complexity alternative,
where  the maximum affordable   complexity is  imposed   through a  proper selection of the VA memory depth $L$.
In  Fig. \ref{vahmsdd}  we compare  the BER  performance of the conventional  hard-decision
GLRT-MSDD
and of the proposed  VA-HMSDD  by setting  the  observation   window size to  $M=7$,  and the VA memory depth to   $L=4$,  5  and 6,  respectively.
It can be seen  from  Fig. \ref{vahmsdd} that
there exists a performance gap between  the conventional hard-decision
GLRT-MSDD and  the proposed  VA-HMSDD when  $L<M$.
The difference in BER  performance  is due   to the different decision metrics employed.
The detection performance of VA-HMSDD is constrained by the memory depth $L$    for a given    $M$,
since $L$ determines  the number of cross-terms in the approximating objective metric function of  (\ref{va}).
By contrast, the conventional hard-decision
GLRT-MSDD  employs the exact objective metric  function of  (\ref{msddmetric2}),
hence    it  achieves   a  better BER performance.
In fact,  VA-HMSDD is based on   an equivalent reformulation of the MSDD criterion
and  it  maintains the BER  performance of the original MSDD  when $M=L$.
Interestingly,  the low-complexity  VA-HMSDD only incurs a small performance loss when  $L$ is  close to $M$, e.g.,
the VA-HMSDD with $L=6$  suffers  a  SNR penalty of
about 0.2$\mathrm{dB}$   compared to the conventional hard-decision  GLRT-MSDD with $M=7$.
When the   design parameter $L$   is increased, the detection performance of VA-HMSDD   becomes   closer to that of the original  GLRT-MSDD.
As expected,    at the  $\mathrm{BER} =10^{-4} $,   compared with  the  case of $L = 6$,
VA-HMSDD with $L = 3$   suffers    a BER performance degradation of  about   1$\mathrm{dB}$.
It is worth noting that, when $M$ is large, the conventional hard-decision
GLRT-MSDD  is not
applicable due to its unaffordable complexity. Yet,
our VA-HMSDD  can remain computationally
practical  when selecting a small  $L$,
despite  a little    BER performance degradation.
Moreover,  we consider the BER  performance comparison of the  BP-MSDD and the  VA-HMSDD  when  their  complexity   are  approximately equal. Fig. \ref{vahmsdd}  demonstrates that when $M_{1}=5$ for BP-MSDD, $M_{2}=7$, $L=4$, 5 for VA-HMSDD, their complexity are close to each other.
In this case, the BER  performance  of the BP-MSDD is slightly better than the VA-HMSDD.

\emph{Test Scenario  3}:
In this test scenario,  the  BER performance of the  iterative  BP-MSDD, the iterative  VA-SMSDD
and  the   coded GLRT-MSDD without iteration
are   studied   to  reveal    relevant    insights into the  IDD  receiver design relying on the MSDD philosophy.
In particular,  the  SISO BP-MSDD   is specified  by   (\ref{zui}), and
the low-complexity  VA-SMSDD  is based on the   decision metric  of   (\ref{soft}).
For fair comparison, all  the  receivers  employ a rate-1/2 convolutional code \cite{design}, whose generator polynomial is (133, 171) in octal form and the constraint length is 7.
For the  iterative  BP-MSDD and  the iterative  VA-SMSDD,
four   iterations  are performed  between  the  symbol  detector  and the channel  decoder.
It is observed  from Fig. \ref{Fig10IDD}  that  the  BP-MSDD  based     IDD receiver  enjoys an appealing  BER   performance advantage   compared with the coded GLRT-MSDD.
To better understand the effectiveness of the simplified VA criterion,
the BER performance of the  VA-SMSDD  based   IDD receiver  is
 illustrated     in Fig. \ref{Fig10IDD}
  as well.
It can be seen  that there  exists   a small performance gap between  the BP-MSDD  based IDD receiver  with $M=5$ and the  VA-SMSDD  based IDD receiver  with $L=4$. This is because the  BP-MSDD  based IDD  employs an exact  objective metric  function  of   (\ref{zui}), while the
VA-SMSDD  based  IDD  employs   an approximate simplified   function of   (\ref{soft}).

\emph{Test Scenario 4}:
To consolidate the computational complexity analysis provided in Section V,  the computational
complexity  in terms of multiplication operation  of the BP-MSDD,
the soft-decision MSDD in \cite{pisit},  the  hard-decision GLRT-MSDD, the hard-decision SD-MSDD in \cite{llr},  the VA-SMSDD and the VA-HMSDD
is   illustrated  in
Fig. \ref{complexity}    against different   observation window size $M$  and   different VA memory depth  $L$.
It is observed that the  VA-SMSDD  exhibits  a  complexity reduction
 compared  with  the   BP-MSDD,  especially for a large  value of $M$  and  a small  value of $L$.
This is because the   number of trellis states  for the   BP-MSDD increases exponentially with the observation window size $M$,
which inevitably results in  high computational   complexity.
In contrast  to  the BP-MSDD,   the  complexity of  VA-SMSDD  increases exponentially with  the memory depth $L$, which satisfies   $L \leq M $.
Hence, a lower  complexity can be obtained when selecting a small $L$ for the  VA-SMSDD.
It is also  observed from Fig. \ref{complexity}  that   the proposed VA-HMSDD enjoys a complexity reduction  compared to the hard-decision GLRT-MSDD,  regardless of the specific values of $M$;
the complexity of  the soft-decision MSDD in   \cite{pisit}  is lower than the BP-MSDD,
and the complexity of the hard-decision SD-MSDD in \cite{llr} is slightly lower than that of the hard-decision
GLRT-MSDD.

\begin{figure}
\begin{centering}
\includegraphics[width=7.5cm]{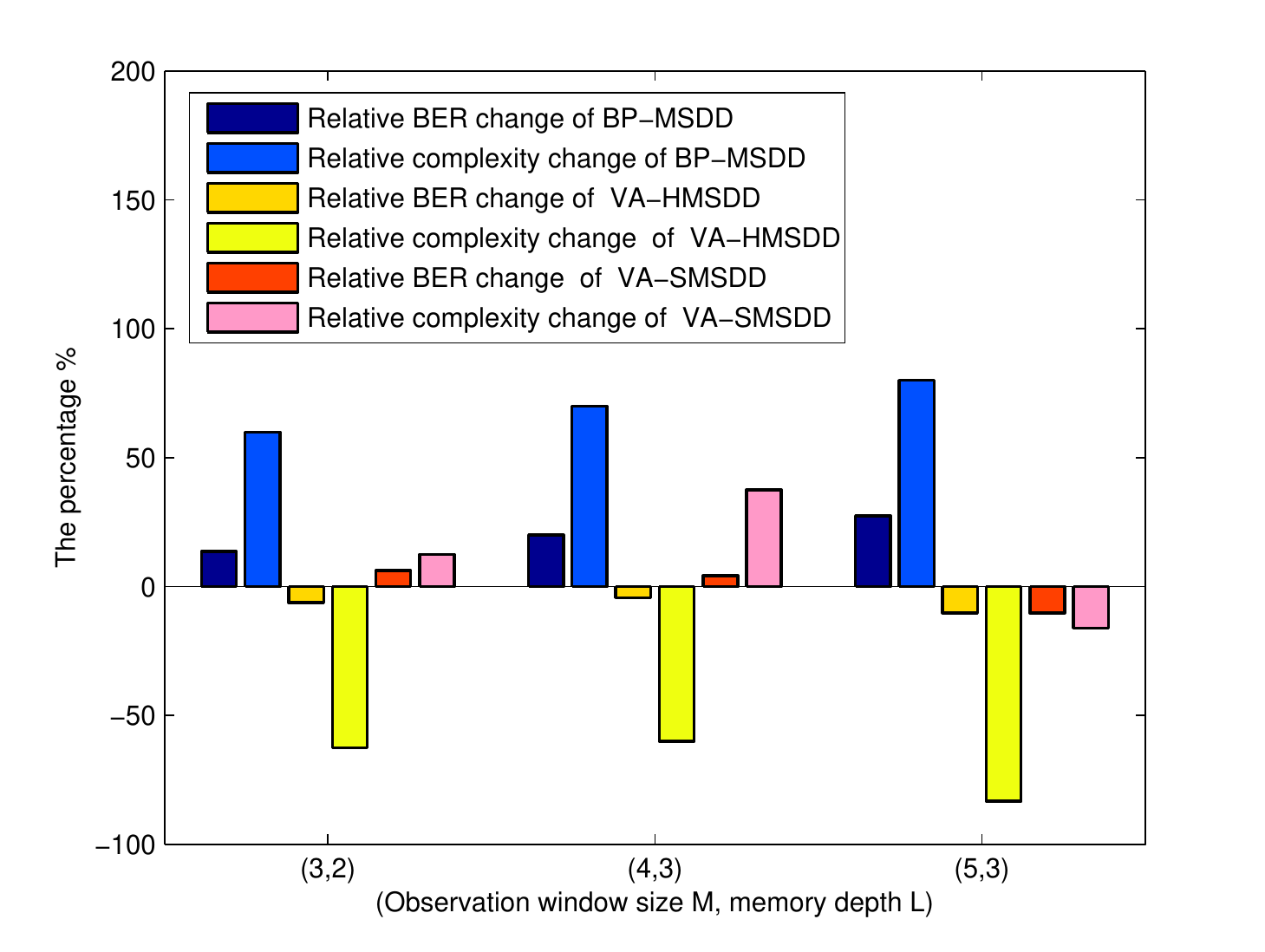}
\caption{\label{BERcomduibi} BER performance and computational  complexity  comparisons   of the BP-MSDD, VA-HMSDD,   VA-SMSDD against the  GLRT-MSDD in terms of relative percentage
with different combinations of  $(M,L)=(3,2)$, $(4,3)$  and $(5,3)$, respectively.  The  relative complexity   and  the relative  BER performance   are  defined as $\frac{C_{\mathrm{CA}}-C_{\mathrm{GLRT}}}{C_{\mathrm{GLRT}}}\times100\% $  and
$\frac{B_{\mathrm{GLRT}}-B_{\mathrm{CA}}}{B_{\mathrm{GLRT}}}\times100\% $, respectively.
}\par
\end{centering}
\end{figure}

In addition,  the relative BER performance and computational complexity results under several different combinations of the parameter values $(M, L)$
  are  illustrated in Fig. 12.
The positive   percentage  represents   the BER performance  improvement and the computational  complexity increase compared to those of  the benchmark  GLRT-MSDD;
meanwhile,  negative  percentage   represents   the BER performance loss and
the computational  complexity reduction compared to those of  the benchmark  GLRT-MSDD.
The  relative complexity   and  the relative  BER performance   are  defined as $\frac{C_{\mathrm{CA}}-C_{\mathrm{GLRT}}}{C_{\mathrm{GLRT}}}\times100\% $  and
 $\frac{B_{\mathrm{GLRT}}-B_{\mathrm{CA}}}{B_{\mathrm{GLRT}}}\times100\% $, respectively,
 where  $C_{\mathrm{CA}}$  and  $B_{\mathrm{CA}}$     represent the complexity and the BER performance of the compared algorithm, respectively;
  while     $C_{\mathrm{GLRT}}$  and  $B_{\mathrm{GLRT}}$   are   the complexity  and the BER performance of the GLRT, respectively.
From   Fig. 12,   we can see  that for all the combinations of $(M,L)=(3,2)$, $(4,3)$  and $(5,3)$,  the BER performance of the  BP-MSDD is superior to that of  the noncoherent GLRT-MSDD  at the cost of   increased computational    complexity;
by contrast, for these combinations of $(M, L)$ values, the  VA-HMSDD  imposes a reduced computational
 complexity  at the expense of  a  small  performance  loss.
Notably,  the   VA-SMSDD shows  both positive and negative percentages under these combinations of (M, L) values, which indicates that the BER performance and the computational complexity of the VA-SMSDD are bounded by those of   the BP-MSDD and the VA-HMSDD.

\section{Conclusions}

A fundamental  BP-MSDD framework   is  proposed   for    efficiently calculating   the  \emph{ a posteriori} information    of the  symbols transmitted.
The   BP-MSDD achieves appealing  performance by exploiting a trellis having a full number of states.
Furthermore,  to facilitate the   practical implementations of the  MSDD,
the proposed  BP-MSDD is reformulated  from the perspective of  LLR,
where the likelihood of  MSDD is  simplified, and  a low-complexity  VA-HMSDD is  derived   with the aid of  a trellis having a reduced number of states.
Additionally,  a soft-decision  based VA-SMSDD is proposed, which achieves better performance than the hard-decision based VA-HMSDD.
Both the BP-MSDD and the VA-SMSDD are capable of flexibly
providing   tradeoffs  between the achievable  performance and the computational  complexity imposed.
Finally,   since both of the   proposed    BP-MSDD and   VA-SMSDD  are SISO  algorithms, they are capable of substantially improving the receiver's BER performance by invoking the IDD philosophy.
Simulation results have demonstrated the effectiveness of the  proposed algorithms  in the illustrative context of DSTBC aided UWB-IR systems.

\section*{Acknowledgement}

The authors would like to thank Dr. Taotao Wang, who helped us improve the manuscript.

\bibliographystyle{IEEEtran}
%\bibliography{OIA}

\begin{thebibliography}{99}

\bibitem{50years}
S. Yang  and L. Hanzo, ``Fifty years of MIMO detection: The road to large-scale MIMOs,'' \emph{IEEE Commun. Surveys Tuts.},
vol. 17, no. 4,  pp. 1941--1988, 4th Quart. 2015.
\bibitem{noncoherentdete}
G. Foschini, L. Greenstein,  and   G. Vannucci, ``Noncoherent detection of coherent lightwave signals corrupted by phase noise,'' \emph{IEEE Trans.  Commun.},
vol. 36, no. 3,  pp. 306--314, Mar. 1998.
\bibitem{energycapture}
M. Win and R. Scholtz, ``On the energy capture of ultrawide bandwidth signals in dense multipath environments,'' \emph{IEEE Commun. Lett.}, vol. 2, no. 9,
pp. 245--247, Sep. 1998.
\bibitem{noncoherent}
M. Simon and M. Alouini, ``A unified approach to the probability of error for noncoherent and differentially coherent modulations over generalized fading channels,'' \emph{IEEE Trans.  Commun.}, vol. 46,  no. 12,  pp. 1625--1638, Dec. 1998.
\bibitem{differential}
H. Leib  and  S. Pasupathy, ``The phase of a vector perturbed by Gaussian noise and differentially coherent receivers,'' \emph{IEEE Trans. Inf. Theory.}, vol. 34,
no. 6, pp. 1491--1501, Nov. 1988.
\bibitem{hop}
P. Ho and D. Fung, ``Error performance of multiple symbol differential detection of PSK signals transmitted over correlated Rayleigh fading channels,'' in \emph{Proc. IEEE Int. Conf. Commun. (ICC 1991)}, Denver, CO, USA, Jun. 1991, pp. 23--26.
\bibitem{fada}
F. Adachi and M. Sawahashi,  ``Decision feedback multiple-symbol differential detection for M-ary DPSK,'' \emph{Electron. Lett.}, vol. 29, no. 15, pp. 1385--1387, Jul. 1993.
\bibitem{aabra}
A. Abrardo, G. Benelli, G. Cau, ``Multiple-symbol differential detection of GMSK for mobile communications,'' \emph{IEEE Trans. Veh. Technol.}, vol. 44, no. 3, pp. 379--390, Aug. 1995.
\bibitem{lampe}
L. Lampe, R. Schober, V. Pauli, and C. Windpassinger, ``Multiple-symbol differential sphere decoding,'' \emph{IEEE Trans. Commun.}, vol. 53, no. 12, pp. 1981--1985, Dec. 2005.
\bibitem{mpsk}
D. Divsalar and M. Simon, ``Multiple symbol differential detection of MPSK,'' \emph{IEEE Trans.  Commun.},  vol. 38, no. 3,   pp. 300--308, Mar. 1990.
\bibitem{tianzhi}
V. Lottici and Z. Tian,  ``Multiple symbol differential detection for UWB communications,'' \emph{IEEE Trans. Wireless Commun.}, vol. 7,  no. 5,
pp. 1656--1666, May 2008.
\bibitem{efficientcoherent}
D. Warrier and U. Madhow, ``Spectrally efficient noncoherent communication,'' \emph{IEEE Trans. Inf. Theory.}, vol. 48,  no. 3,   pp. 651--668, Mar. 2002.
\bibitem{space-time}
V. Tarokh, H. Jafarkhani, and A. R. Calderbank, ``Space-time block codes from orthogonal designs,'' \emph{IEEE Trans. Inf. Theory.}, vol. 45, no. 5,
pp. 1456--1467, Jul. 1999.
\bibitem{4v}
V. Tarokh and H. Jafarkhani, ``A differential detection scheme for transmit diversity,'' \emph{IEEE J. Sel. Areas Commun.}, vol. 18, no. 7, pp. 1169--1174, Jul. 2000.
\bibitem{5p}
P. Fan, ``Multiple-symbol detection for transmit diversity with differential encoding scheme,'' \emph{IEEE Trans. Consum. Electron.}, vol. 47, no. 1, pp. 96--100, Feb. 2001.
\bibitem{wtt}
T. Wang, T. Lv,  H. Gao, and Y. Lu, ``BER analysis of decision-feedback multiple symbol detection in non-coherent MIMO ultra-wideband systems,'' \emph{IEEE Trans. Veh. Technol.}, vol. 62, no. 9,  pp.  4684--4690, Nov.   2013.
\bibitem{block}
T. Wang, T. Lv, and H. Gao, ``Sphere decoding based multiple symbol detection for differential space-time block coded ultra-wideband systems,'' \emph{IEEE Commun. Lett.}, vol. 15, no. 3, pp. 269--271, Mar. 2011.
\bibitem{iterativenon}
R.  Fischer, L.  Lampe, and S.  Weinfurtner, ``Coded modulation for noncoherent reception with application to OFDM,'' \emph{IEEE Trans. Veh. Technol.}, vol. 50, no. 4,  pp. 910--919, Jul. 2001.
\bibitem{pisit}
P. Vanichchanunt, P. Sangwongngam, S. Nakpeerayuth, and L. Wuttisittikulkij, ``Iterative multiple symbol differential detection for Turbo coded differential unitary space-time modulation,'' \emph{Journal of Communication and Networks}, vol. 10, no. 1,  pp. 44--54, Mar. 2008.
\bibitem{mars}
I. Marsland and  P. Mathiopoulos,  ``On the performance of iterative noncoherent detection of coded MPSK signals,'' \emph{IEEE Trans.  Commun.}, vol. 48, no. 4,  pp. 588--596,  Apr. 2000.
\bibitem{llr}
V. Pauli, L. Lampe, and R. Schober, ````Turbo DPSK'' using soft multiple symbol differential sphere decoding,'' \emph{IEEE Trans. Inf. Theory.}, vol. 52,  no. 4,  pp. 1385--1398, Apr. 2006.
\bibitem{con}
B.  Hochwald and S. ten Brink, ``Achieving near-capacity on a multiple-antenna channel,'' \emph{IEEE Trans.  Commun.}, vol. 51, no. 3,  pp. 389--399, Mar. 2003.
\bibitem{Turbo}
C. Berrou, A. Glavieux, and P. Thitimajshima, ``Near Shannon limit error-correcting coding and decoding: Turbo Codes,'' in \emph{Proc. IEEE Int. Conf. Commun. (ICC 1993)}, Geneva, Switzerland, May 1993, pp. 1064-1070.
\bibitem{yshaoshi}
S. Yang, L. Wang, T. Lv,  and L. Hanzo, ``Approximate Bayesian probabilistic-data-association-aided iterative detection for MIMO systems using arbitrary M-ary modulation,'' \emph{IEEE Trans. Veh. Technol.}, vol. 62, no. 3,  pp. 1228--1240, Mar. 2013.
\bibitem{yshaoshitcom}
S. Yang, T. Lv, R.  Maunder, and L. Hanzo, ``From nominal to true a posteriori probabilities: An exact Bayesian theorem based probabilistic data association approach for iterative MIMO detection and decoding,'' \emph{IEEE Trans. Commun.}, vol. 61, no. 7,  pp. 2782--2793, Jul.  2013.
\bibitem{frfacotr2001}
F.  Kschischang, B.  Frey, and H. Loeliger, ``Factor graphs and the sum-product algorithm,'' \emph{IEEE Trans. Inf. Theory.},  vol. 47, no. 2,  pp. 498--519,
Feb. 2001.
\bibitem{design}
M.  Fossorier, S. Lin, and D. Costello, ``On the weight  distribution of terminated convolutional codes,'' \emph{IEEE Trans. Inf. Theory.}, vol. 45, no. 5, pp. 1646--1648, Jul. 1999.
\bibitem{BP2}
M. Fossorier, M. Mihaljevic, and H. Imai, ``Reduced complexity iterative  decoding of low density parity check codes based on belief propagation,'' \emph{IEEE Trans.  Commun.},  vol. 47, no. 5, pp. 673--680, May 1999.
\bibitem{1974}
L. Bahl, J. Cocke, F. Jelinek, and J. Raviv, ``Optimal decoding of linear codes for minimizing symbol error rate,'' \emph{IEEE Trans. Inf. Theory.}, vol. 20, no. 2, pp. 284--287, Mar. 1974.
\bibitem{arxiv}
T. Wang, T. Lv,  H. Gao, and S. Zhang  (2015, Apr. 5). ``Joint multiple symbol differential detection and channel decoding for noncoherent UWB impulse radio by belief propagation,'' [Online]. Available: http://arxiv.org/abs/1504.01100.
\bibitem{factor2004}
H. Loeliger. ``An introduction to factor graphs,'' \emph{IEEE Signal Processing Mag.}, vol. 21, no. 1, pp. 28--41, Jan. 2004.
\bibitem{decoding}
H. Niu, M. Shen, J. A. Ritcey, and H. Liu, ``A factor graph approach to iterative channel estimation and LDPC decoding over fading channels,'' \emph{IEEE Trans. Wireless Commun.}, vol. 4, no. 4,  pp. 1345-1350, Jul.  2005.
\bibitem{win}
M. Z. Win and R. A. Scholtz, ``Ultra-wide bandwidth time-hopping spread-spectrum impulse radio for wireless multiple-access communications,'' \emph{IEEE Trans.  Commun.}, vol. 48,  no. 4,  pp. 679--689, Apr. 2000.
\bibitem{zq}
Q. Zhang and C. Ng, ``DSTBC impulse radios with autocorrelation receiver in ISI-free UWB channels,'' \emph{IEEE Trans. Wireless Commun.}, vol. 7, no. 3,  pp. 806--811, Mar. 2008.
\bibitem{jsac}
T. Quek and M. Win, ``Analysis of UWB transmitted-reference communication systems in dense multipath channels,''  \emph{IEEE J. Sel. Areas Commun.}, vol. 23, no. 9, pp. 1863--1874, Sep. 2005.
\bibitem{bys}
Y. S. Shmaliy,  ``On the multivariate conditional probability density of a vector perturbed by Gaussian noise,'' \emph{IEEE Trans. Inf. Theory.}, vol. 53, no. 12,
pp. 4792--4797, Dec. 2007.
\bibitem{memory}
R. W. Chang and J. C. Hancock, ``On receiver structures for channels having memory,'' \emph{IEEE Trans. Inf. Theory.},  vol. 12, no. 4,  pp. 463--468, Oct. 1966.
\bibitem{1order}
C. Tan and N. Beaulieu, ``On first-order Markov modeling for the Rayleigh fading channel,'' \emph{IEEE Trans.  Commun.}, vol. 48,  no. 12, pp. 2032--2040, Dec. 2000.
\bibitem{hidden}
Y. Ephraim and N. Merhav, ``Hidden Markov processes,'' \emph{IEEE Trans. Inf. Theory.},  vol. 48, no. 6,  pp. 1518--1569, Jun. 2002.
\bibitem{mlstc}
Y. Li, C. N. Georghiades, and G. Huang,  ``Iterative maximum-likelihood sequence estimation for space-time coded systems,'' \emph{IEEE Trans.  Commun.},
vol. 49, no. 6, pp. 948--951, Jun. 2001.
\bibitem{lowcom}
G. Rajan and B. Rajan,  ``Algebraic distributed differential space-time codes with low decoding complexity,'' \emph{IEEE Trans. Wireless Commun.}, vol. 7, no. 10,  pp. 3962--3971, Oct. 2008.
\bibitem{Rake}
L. Yang and G. Giannakis, ``Analog space-time coding for multiantenna ultra-wideband transmissions,'' \emph{IEEE Trans. Commun.}, vol. 52, no. 3, pp. 507--517, Mar. 2004.
\end{thebibliography}

\begin{IEEEbiography}[{\includegraphics[width=1in,height=1.25in,clip,keepaspectratio]{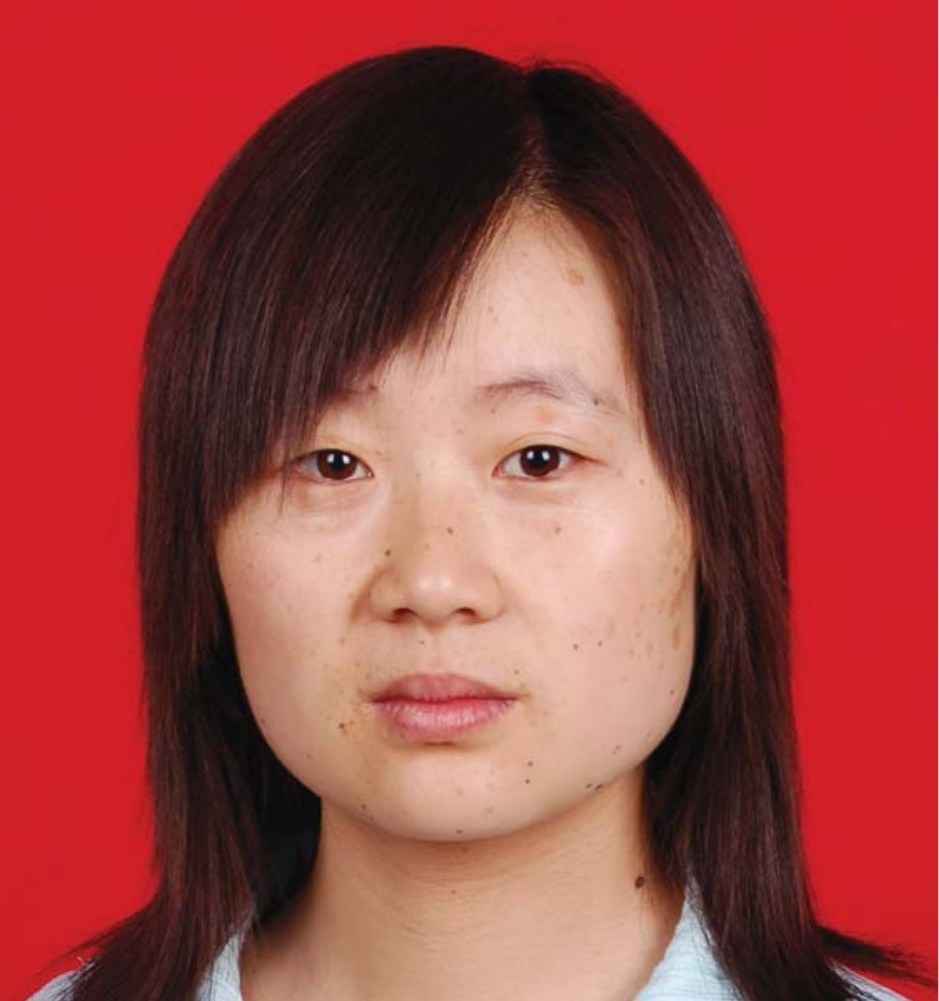}}]{Chanfei Wang}
 is now working towards the Ph.D. degree in information engineering from Beijing University of Posts and Telecommunications (BUPT), Beijing, China. Her research interests focus on signal processing techniques in ultra-wideband wireless communications.
\end{IEEEbiography}
\begin{IEEEbiography}[{\includegraphics[width=1in,height=1.25in,clip,keepaspectratio]{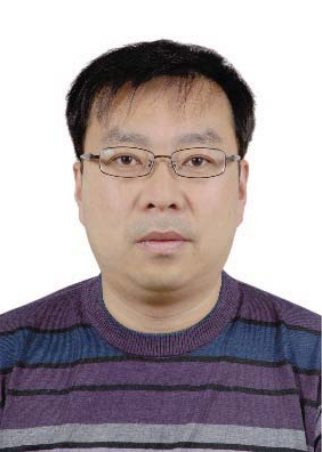}}]{Tiejun Lv}
(M'08-SM'12) received the M.S. and Ph.D. degrees in electronic engineering from the University of Electronic Science and Technology of China (UESTC), Chengdu, China, in 1997 and 2000, respectively. From January 2001 to January 2003, he was a Postdoctoral Fellow with Tsinghua University, Beijing, China. In 2005, he became a Full Professor with the School of Information and Communication Engineering, Beijing University of Posts and Telecommunications (BUPT). From September 2008 to March 2009, he was a Visiting Professor with the Department of Electrical Engineering, Stanford University, Stanford, CA, USA. He is the author of more than 200 published technical papers on the physical layer of wireless mobile communications. His current research interests include signal processing, communications theory and networking. Dr. Lv is also a Senior Member of the Chinese Electronics Association. He was the recipient of the Program for New Century Excellent Talents in University Award from the Ministry of Education, China, in 2006. He received the Nature Science Award in the Ministry of Education of China for the hierarchical cooperative communication theory and technologies in 2015.
\end{IEEEbiography}
\begin{IEEEbiography}[{\includegraphics[width=1in,height=1.25in,clip,keepaspectratio]{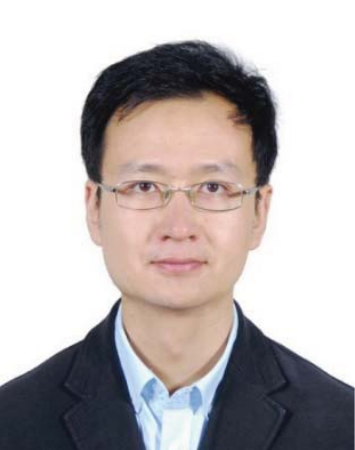}}]{Hui Gao}
(S'10-M'13-SM'16) received his B. Eng. degree in Information Engineering and Ph.D. degree in Signal and Information Processing from Beijing University of Posts and Telecommunications (BUPT), Beijing, China, in July 2007 and July 2012, respectively. From May 2009 to June 2012, he also served as a research assistant for the Wireless and Mobile Communications Technology R and D Center, Tsinghua University, Beijing, China. From Apr. 2012 to June 2012, he visited Singapore University of Technology and Design (SUTD), Singapore, as a research assistant. From July 2012 to Feb. 2014, he was a Postdoc Researcher with SUTD. He is now with the School of Information and Communication Engineering, Beijing University of Posts and Telecommunications (BUPT), as an assistant professor. His research interests include massive MIMO systems, cooperative communications, ultra-wideband wireless communications.
\end{IEEEbiography}
\begin{IEEEbiography}[{\includegraphics[width=1in,height=1.25in,clip,keepaspectratio]{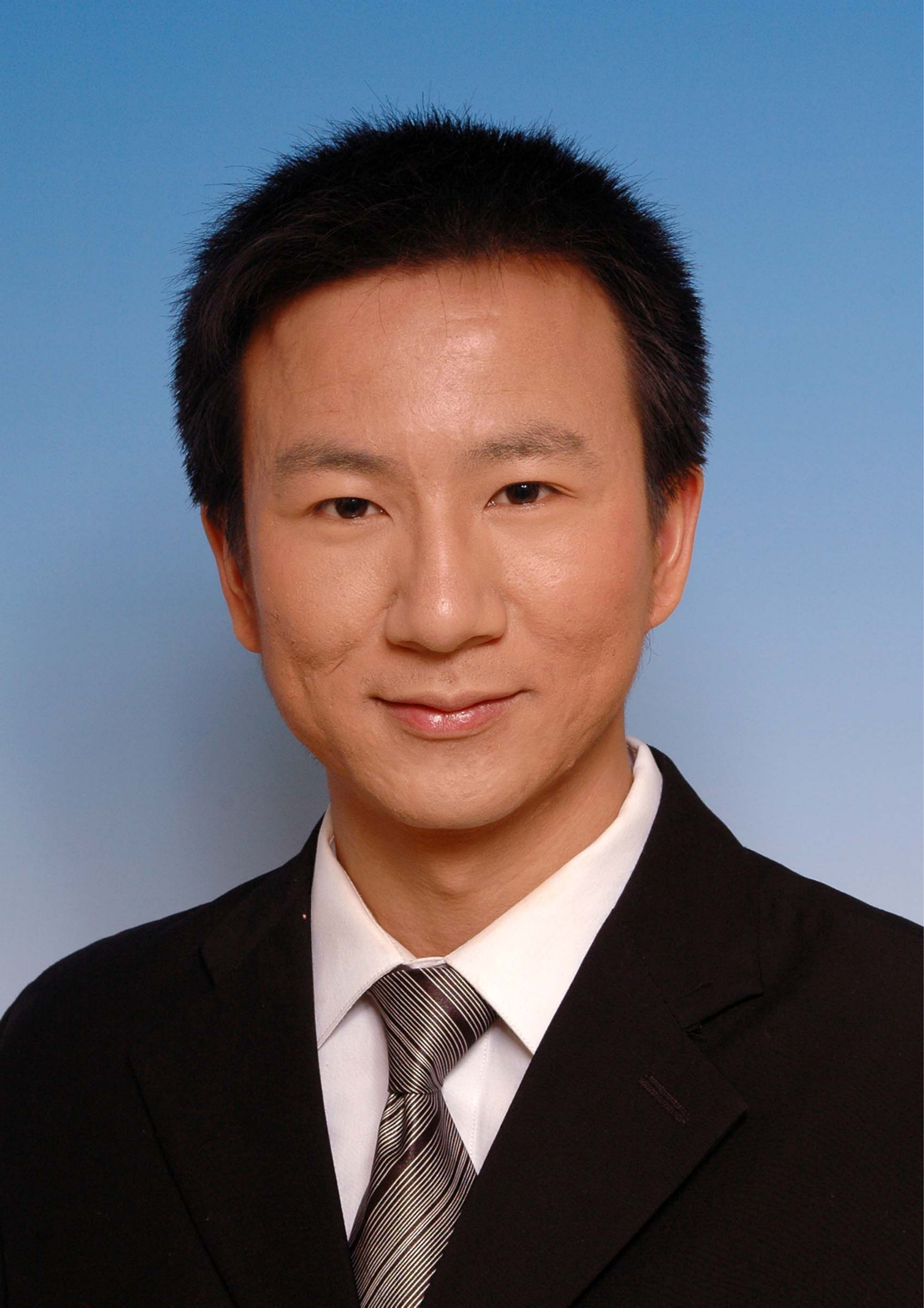}}]{Shaoshi Yang}
(S'09-M'13) received his B.Eng. degree in Information Engineering from Beijing University of Posts and Telecommunications (BUPT), Beijing, China in Jul. 2006, his first Ph.D. degree in Electronics and Electrical Engineering from University of Southampton, U.K. in Dec. 2013, and his second Ph.D. degree in Signal and Information Processing from BUPT in Mar. 2014. He is now working as a Postdoctoral Research Fellow in University of Southampton, U.K. From November 2008 to February 2009, he was an Intern Research Fellow with the Communications Technology Lab (CTL), Intel Labs, Beijing, China, where he focused on Channel Quality Indicator Channel (CQICH) design for mobile WiMAX (802.16m) standard. His research interests include MIMO signal processing, green radio, heterogeneous networks, cross-layer interference management, convex optimization and its applications. He has published in excess of 30 research papers on IEEE journals. Shaoshi has received a number of academic and research awards, including the prestigious Dean's Award for Early Career Research Excellence at University of Southampton, the PMC-Sierra Telecommunications Technology Paper Award at BUPT, the Electronics and Computer Science (ECS) Scholarship of University of Southampton, and the Best PhD Thesis Award of BUPT. He is a member of IEEE/IET, and a junior member of Isaac Newton Institute for Mathematical Sciences, Cambridge University, U.K. He also serves as a TPC member of several major IEEE conferences, including IEEE ICC, GLOBECOM, VTC, WCNC, PIMRC, ICCVE, HPCC, and as a Guest Associate Editor of IEEE Journal on Selected Areas in Communications. (http://sites.google.com/site/shaoshiyang/).

\end{IEEEbiography}

\end{document}